\documentclass[preprint,5p,times,twocolumn]{elsarticle}

\usepackage{amsmath,amssymb,amsfonts}
\renewcommand{\baselinestretch}{0.98}
\usepackage[inline]{enumitem}
\usepackage{graphicx} 
\usepackage{textcomp} 
\usepackage{comment}
\specialcomment{new}{\begingroup\color{blue}}{\endgroup}
\specialcomment{red}{\begingroup\color{red}}{\endgroup}
\usepackage{array}
\newcolumntype{P}[1]{>{\centering\arraybackslash}p{#1}}
\usepackage{listings}
\usepackage{algpseudocode}
\usepackage{csquotes}
\usepackage{rotating}
\usepackage{wasysym}
\usepackage[table]{xcolor}
\usepackage{eso-pic}
\usepackage{multirow}
\usepackage{float} 
\usepackage{subfigure}
\usepackage{booktabs}
\usepackage{url}
\usepackage{soul}
\usepackage{enumitem}
\newlist{steps}{enumerate}{1}
\setlist[steps, 1]{label = S.\arabic*:}
\usepackage{longtable}
\usepackage{multirow,tabularx}
\newcolumntype{Y}{>{\centering\arraybackslash}X}
\usepackage{balance}

\usepackage{algpseudocode}
\usepackage[linesnumbered,ruled]{algorithm2e}
\usepackage{algpseudocode}
\usepackage{graphicx} 
\usepackage{float} 
\usepackage{textcomp} 
\usepackage{eso-pic}
\usepackage{xcolor}
\usepackage{comment}
\usepackage{lipsum}
\usepackage{mathtools}
\usepackage{cuted}
\usepackage{csquotes}



\journal{Arxiv}

\begin{document}

\begin{frontmatter}


\title{Blockchain-based AI Methods for Managing Industrial IoT: Recent Developments, Integration Challenges and Opportunities}

\author[AR]{Anichur Rahman
}
\ead{anis\_cse@niter.edu.bd}
\author[AR]{Dipanjali Kundu} \ead{dkundu@niter.edu.bd}
\author[TD]{Tanoy Debnath} \ead{tanoy.debnath@stonybrook.edu}
\author[AR]{Muaz Rahman} \ead{muaz@niter.edu.bd}

\author[JI]{Md. Jahidul Islam} \ead{jahid@cse.green.edu.bd}


\address[AR]{Department of Computer Science and Engineering, National Institute of Textile Engineering and Research (NITER), \\Constituent Institute of the University of Dhaka, Savar, Dhaka-1350, Bangladesh}

\address[TD]{Department of Computer Science, Stony Brook University, Stony Brook, NY, USA}


\address[JI]{Department of Computer Science and Engineering, Green University of Bangladesh, Kanchon, Dhaka, Bangladesh}




\begin{abstract}
Currently, Blockchain (BC), Artificial Intelligence (AI), and smart Industrial Internet of Things (IIoT) are not only leading promising technologies in the world, but also these technologies facilitate the current society to develop the standard of living and make it easier for users. However, these technologies have been applied in various domains for different purposes. Then, these are successfully assisted in developing the desired system, such as--smart cities, homes, manufacturers, education, and industries. Moreover, these technologies need to consider various issues--security, privacy, confidentiality, scalability, and application challenges in diverse fields. In this context, with the increasing demand for these issues solutions, the authors present a comprehensive survey on the AI approaches with BC in the smart IIoT. Firstly, we focus on state-of-the-art overviews regarding AI, BC, and smart IoT applications. Then, we provide the benefits of integrating these technologies and discuss the established methods, tools, and strategies efficiently. Most importantly, we highlight the various issues--security, stability, scalability, and confidentiality and guide the way of addressing strategy and methods. Furthermore, the individual and collaborative benefits of applications have been discussed. Lastly, we are extensively concerned about the open research challenges and potential future guidelines based on BC-based AI approaches in the intelligent IIoT system. 
\end{abstract}
\vspace{2mm}


\end{frontmatter}


\section{Introduction}
\textcolor{black}{Industry 4.0 \cite{dhirani2020hybrid} is a novel and smart concept of an extensively connected factory that implements innovative technologies (such as IoT, cloud computing, artificial intelligence.) and state-of-the-art solutions (i.e., IIoT, automation, monitoring, etc.) to improve the manufacturing conditions with reduced expenses, dexterity, improved efficiency, and performing remote activities, among other things. Data and network security are critical in such an autonomous system. At a glance, additive manufacturing techniques, autonomous and collaborative robotics, the Industrial Internet of Things (IIoT), big data analytics, and cloud manufacturing processes are the main technologies that enable 4IR to be sustained \cite{becue2021artificial}. The current scenarios demonstrate the benefits of IIoT in improving QoS in industries, starting with predictive maintenance and progressing to remote asset control and the deployment of the Digital Twin concept \cite{udoy20234sqr}, which allows the owner to virtualize the operations environment and be proactive when any anomalies are detected. Despite the fact that IIoT brings value to traditional industries, a balance must be struck between operational benefits and security levels. However, the applications of the IIoT can be divided into four groups. Here, production flow, quality control, and energy usage are all part of the first class. The goal of this lesson is to improve industrial processes. Management, which is concerned with operations, including supply chain and corporate decision, is the second type of application. The third class is concerned with resource allocation and collaboration. Technology for collaborative production and customization is covered in this lesson. Moreover, the final lesson is mostly concerned with the management of product lifetime. Also, it emphasizes performance enhancements, which include remote monitoring and the ability to trace products \cite{yu2019toward}. As the number of IIoT applications and services grows, so does the range of cyber security threats, necessitating the development of new security measures and controls \cite{kim20205g,farkas20195g}.}
\vspace{1.5mm}

\textcolor{black}{On the other hand, in recent years, the use of Artificial Intelligence (AI) in IIoT has gained massive interest, particularly due to the effectiveness of deep learning in addressing previously regarded difficult concerns, such as security and privacy. AI is being tested in a wide range of environments, which includes healthcare, finance, and driverless cars. Along with AI, other technologies such as Internet-of-Things (IoT) have boosted other fields such as Industry 4.0, where corporations seek to have smarter manufacturing that can be tailored to their customer's needs through the integration of Industrial-IoT (IIoT) into the production chain. Moreover, smart machines, equipment, and integrated automated systems \cite{ashima2021automation} can now automate ordinary processes and handle complicated issues, reducing the requirement for human intervention. Intelligent workspace enhancements, smart data exploration, and rationale automation, together with other facets of corporate smartness, need to be considered.
Furthermore, M2M communication infrastructure is an essential component of interconnected industries, with an increased reliance on cutting-edge wireless communication systems (5G, time-sensitive networking, etc.) \cite{farkas20195g} coupled with automated, self-driven, and improved network characteristics. In the future, M2M devices are expected to function autonomously and relay decisions depending on AI and ML algorithms. Such devices require minimal human involvement and a high degree of security because they possess the ability to impose disrupting hazards \cite{bartoli2015advanced, Islam2021}. However, the principal area of potential threats for M2M communications is unexpected internet breaches, software that is mostly discovered after the fact, reduced abilities as a result of low energy, economic barriers, distant locations, and legacy systems.}
\vspace{1mm}

\textcolor{black}{At the present world, there is an exponential growth in the availability of internet-connected devices with the development of the IIoT, generating a significant quantity of data. Contrary to traditional IoT networks, data volume and characteristics are important in the IIoT.
As a result, huge amount of data produced by industries necessitates simultaneous smart processing \cite{liang2020toward}. Deep Learning (DL) has the potential to perform a significant part in evaluating Big Data from IIoT networks to gain smart processing ability. Because the technologies concerned with improving the IIoT differ greatly from previous technologies. Moreover, The IIoT has also been referred to as a second industrial revolution. Human working conditions and everyday lifestyles will be drastically altered by emerging IIoT technology \cite{khalil2020network}. Significant amount of current studies have identified IIoT security and confidentiality issues from various perspectives. For example, the authors of \cite{yu2019survey} classified IoT and IIoT security challenges. The research work also stated if these issues apply to IoT, IIoT, or both.
Furthermore, the work highlights the security concerns in the IIoT and emphasizes the importance of developing feasible solutions. Furthermore, the study demonstrates that different IIoT environment necessitate requirement-specific designs. The authors, however, do not propose solutions to the problem of appropriate designs.
The authors of \cite{jayalaxmi2021taxonomy} examined the security issues of IIoT and detailed an overall analysis of potential solutions. The purpose of this study was to define certain open research subjects in the areas of system assimilation, device communication, energy issues, defensive and identification measures, authorization, and IIoT infrastructure. However, the study makes no recommendations for feasible and practical solutions. In a similar vein, Tange et al. \cite{tange2020systematic} conducted a thorough review of the literature on the necessities of IIoT security. The researchers showed the use of fog computing to meet these requirements. Furthermore, the authors detailed certain research opportunities for using protected fog computing in IIoT.
 Similarly, business resources may be jeopardized, resulting in financial consequences \cite{9183413}. As a result, the reliability, fidelity, and threat of IIoT application-failure standards should be greater as compared to IoT applications. Furthermore, IIoT approaches must maintain strict requirements such as simultaneous processing and response, time synchronization, and consistent communication \cite{bader2021searchable}. The CIA triangle refers to the three essential components of security: confidentiality, integrity, and availability. Confidentiality ensures that only certified users can read network data. The consistency of the data ensures that no alterations are performed, and accessibility ensures that all resources and information are available \cite{wan2019blockchain,hasan2021normalized}. On the other hand, data accessibility and consistency are regarded as carrying greater importance than confidentiality in an industrial setting. Confidentiality still carries significant importance in such environment. The attributes discussed above should be raised up to an adequate level using IIoT internet-connected systems. As a result, when developing new IIoT and Industry 4.0 systems, privacy and consistency should be prioritized alongside accessibility \cite{tange2020systematic}.}
\vspace{2mm}
\textcolor{black}{\newline However, deploying AI requires data science expertise, which adds complexity to an already complex environment. While engineers are adept at evaluating massive amounts of data, setting up and establishing production-grade machine learning systems is a difficult task. Unlocking the value of industrial data using artificial intelligence necessitates a hybrid approach. This brings us to the Industrial AI paradigm, which integrates data science and AI with software and domain expertise to offer measurable business outcomes for capital-intensive enterprises. This manufacturing digital nervous system was once mostly reliant on outdated technologies and architectures. However, the sector is already adopting from IT methodologies and architectures designed for enterprise systems, with an emphasis on speed and size. These next-generation Industrial AI solutions make it easy for the industry to embed and deploy AI into industrial systems, allowing users to address industrial challenges without having to retrain or add data science capabilities to industrial organizations.}
\vspace{2mm}

\textcolor{black}{\newline Recent IIoT surveys have concentrated on the general IoT domain rather than the IIoT domain. They either provided a broad overview of IoT security \cite{sequeiros2020attack} or a detailed security analysis focused on specific IoT technologies or a specific layer of IoT architecture \cite{tabrizi2019design}. Furthermore, several surveys \cite{fernandez2018review, hassija2019survey, amanullah2020deep, lao2020survey} investigated the relationship between IoT security and Blockchain technologies. In the IIoT domain, survey directions have recently been directed to be hammered down \cite{polychronou2021comprehensive}. These IIoT security surveys are focused on deep learning in IIoT threat detection \cite{wu2021deep}, and decentralized BC technologies \cite{latif2021blockchain}. However, none of them conducts a thorough security analysis of IIoT architecture and recent industry solutions.
It should be noted that the world is entering a new era, one of globalization and digital technology evolution. This evolution is causing changes in vertical sectors as well as having a direct impact on society. The environment and citizens' values are becoming more diverse and complex. This digital transformation is quickly becoming a cornerstone of industrial strategy \cite{borawake2020green}. This transformation has completely altered the mode of operation of the vast majority of industrial applications and businesses. Employees capable of understanding, implementing, and developing new work models are now required by the latter. As a result, the stakes of this transformation are defined by process digitization mastery, business line redefinition, good data analysis, and agile integration within the company.}
\vspace{2mm}
\textcolor{black}{\newline On the other hand, Many businesses use AI algorithms to make real-time choices in their IIoT applications. When it comes to AI-based solutions, it is critical to recognize that data is king. The most significant part of applying AI for optimizing an organization and obtaining insights is aggregating, cleansing, and preparing unique data. Before AI developers can start training their machine learning models, they generally spend up to 75 percent of their time just digesting the raw data. Remember that in order to train a machine learning model running on IIoT devices, you'll need a data set or series of data sets that reflect the actual conditions and situations you'll be working with when the application is live. However, just because of multiple computing and storage problems, existing solutions are unable to compete with human decision-making capabilities. If IoT devices are automated, they can make decisions for themselves. Only when all activities are adequately coordinated can a factory work normally. Even in a single sector or part of the assembly line, any failure might stymie the entire operation and put production on hold indefinitely. Any irregular changes in data in smart systems can signal a problem, and devices in succeeding sections can be delayed to preserve resources \cite{zhang2019knowledge}. 
Furthermore, innovative industries collect a large amount of data from a variety of sources and use it to create useful business solutions and insights. The benefits of implementing autonomous systems in industries include increased productivity, scalability, cost savings, and flexibility. Furthermore, employing pattern-based and repeated decision-making models intelligent decision-making models in numerous industries may contribute to controlling things like subordinates. Motivated to take this step, this survey research have looked at the practical considerations of integrating security and privacy solutions into smart IIoT system architectures that move away from the cloud paradigm to reduce exposure to threats.}

\begin{table}[h!]
\centering
\caption{List of common abbreviations with description}
 \label{tab:abbrtab}

\begin{tabular}{|p{2.0cm}|p{5.5cm}|}
 \toprule
\textbf{Keys} & \textbf{Description}\\
 \midrule
AI & Artificial Intelligence\\
\hline
ANN & Artificial Neural Networks\\
\hline
BC & Blockchain\\
\hline
CIA & Confidentiality, Integrity, and Availability\\
\hline
COVID-19 & Coronavirus Disease 2019\\
\hline
DAC & Distributed Autonomous Corporations\\
\hline
DoS & Denial of Service\\
\hline
DDoS & Distributed Denial of Service\\
\hline
DL & Deep Learning\\
\hline
DT & Decision Tree\\
\hline
FL & Federated Learning\\
\hline
GA & Genetic Algorithm\\
\hline
IoT & Internet of Things\\
\hline
IIoT & Industrial Internet of Things\\
\hline
IP & Internet Protocol\\
\hline
LPU & Local Processing Unit\\
\hline
MCDM & Multi-criteria Decision Making \\
\hline
ML & Machine Learning\\
\hline
M2M & Machine to Machine\\
\hline
MIMO & Multiple Input, Multiple Output\\
\hline
NLP & Natural Language Processing\\
\hline
P2P & Peer-to-Peer\\
\hline
PCA & Patient Centric Agent\\
\hline
PDA & Personal Digital Assistants\\
\hline
PM & Patient Management\\
\hline
QoS & Quality of Services\\
\hline
SH & Smart Healthcare\\
\hline
SC & Smart Contact\\
\hline
SDN & Software Defined Networking\\
\hline
SDP & Sensor Data Provider\\
\hline
SVM & Support Vector Machine\\
\hline
WIoT & Wireless Internet of Things\\
\hline
WSN & Wireless Sensor Networks\\
\hline
DAO & Decentralized Autonomous Organization\\
\hline
OT & Operational Technology\\
\hline
\end{tabular}
\end{table}

\begin{table*}
\scriptsize

\caption{\textcolor{black}{Related surveys regarding BC, AI, and IIoT. The works are grouped, based on the related technology, and reported chronologically}}\label{tab:T1}
\centering
\begin{tabular}{|p{2.0 cm}|p{0.5 cm}|p{1.3 cm}|p{12.0 cm}|}
\hline

\textbf{Works} & \textbf{Year} & \textbf{Used Technology} & \textbf{Major Contributions} \\  
\hline
Ali et al.~\cite{ali2019blockchain} & 2019 & \multirow{10.5}{*}{\textcolor{black}{Blockchain}} & Blockchain-based smart-IoT trust zone measurement architecture. \\
\cline{1-1} \cline{2-2} \cline{4-4}

Moniruzzaman et al.~\cite{moniruzzaman2020blockchain} & 2020 & & Blockchain for smart homes: Review of current trends and research challenges. \\
\cline{1-1} \cline{2-2} \cline{4-4}

Shrestha et al.~\cite{shrestha2020blockchain} & 2020 & & A Blockchain Platform for User Data Sharing Ensuring User Control and Incentives.\\
\cline{1-1} \cline{2-2} \cline{4-4}

Hou et al.~\cite{hou2020iot} & 2020 & & IoT and Blockchain-based smart agri-food supply chains.\\
\cline{1-1} \cline{2-2} \cline{4-4}

Majeed et al.~\cite{majeed2021blockchain} & 2021 & & Comprehensively reviewed the role of Blockchain in enabling IoT-based smart cities.\\
\cline{1-1} \cline{2-2} \cline{4-4}

Ahmed et al.~\cite{abd2021quantum} & 2021 & & Quantum-Inspired Blockchain-Based Cybersecurity: Securing Smart Edge Utilities in IoT-Based Smart Cities.\\
\cline{1-1} \cline{2-2} \cline{4-4}

Sahal et al.~\cite{sahal2022blockchain} & 2022 & & Digital Twins Collaboration for Smart Pandemic Alerting.\\
\cline{1-1} \cline{2-2} \cline{4-4}

Hannah et al.~\cite{hannah2022blockchain} & 2022 & & Blockchain-based Deep Learning to Process IoT Data Acquisition in Cognitive Data.\\
\cline{1-1} \cline{2-2} \cline{4-4}

Abdelzahir et al.~\cite{abdelmaboud2022blockchain} & 2022 & & Blockchain for IoT Applications: taxonomy, platforms, recent advances, challenges and future research directions.\\
\cline{1-1} \cline{2-2} \cline{4-4}

Lin et al.~\cite{lin2022survey} & 2022 & & Research directions on Blockchain smart contract.\\
\cline{1-1} \cline{2-2} \cline{3-3} \cline{4-4}

Valanarasu et al.~\cite{valanarasu2019smart} & 2019 & \multirow{11.5}{*}{AI} & Smart and secure IoT and AI integration framework for hospital environment.\\
\cline{1-1} \cline{2-2} \cline{4-4}

Ullah et al.~\cite{ullah2020applications} & 2020 & & Applications of artificial intelligence and machine learning in smart cities.\\
\cline{1-1} \cline{2-2} \cline{4-4}

Alshehri et al.~\cite{alshehri2020comprehensive} & 2020 & & A comprehensive survey of the Internet of Things (IoT) and AI-based smart healthcare.\\
\cline{1-1} \cline{2-2} \cline{4-4}

Singh et al.~\cite{singh2020convergence} & 2020 & & Convergence of Blockchain and artificial intelligence in IoT network for the sustainable smart city.\\
\cline{1-1} \cline{2-2} \cline{4-4}

Molinara et al.~\cite{molinara2021artificial} & 2021 & & Highlighted the advancement of AI in three areas that are: Smart Cities, Smart Industries, Smart Healthcare.\\
\cline{1-1} \cline{2-2} \cline{4-4}

Lv et al.~\cite{lv2021ai} & 2021 & & AI-empowered IoT security for smart cities.\\
\cline{1-1} \cline{2-2} \cline{4-4}

Sarker et al.~\cite{sarker2022ai} & 2022 & & AI-based modeling: Techniques, applications and research issues towards automation, intelligent and smart systems. \\
\cline{1-1} \cline{2-2} \cline{4-4}

Gupta et al.~\cite{gupta2022artificial} & 2022 & & Artificial intelligence empowered emails classifier for Internet of Things based systems in industry 4.0. \\
\cline{1-1} \cline{2-2} \cline{4-4}

Chaudhary et al.~\cite{chaudhary2022towards} & 2022 & & 5th Generation AI and IoT Driven Sustainable Intelligent Sensors Based on 2D MXenes and Borophene.\\
\cline{1-1} \cline{2-2} \cline{4-4}

Xu et al.~\cite{xu2019explainable} & 2022 & & Explainable AI: A brief survey on history, research areas, approaches and challenges.\\
\cline{1-1} \cline{2-2} \cline{3-3} \cline{4-4}

Vitturi et al.~\cite{vitturi2019industrial}  & 2019 & \multirow{13.5}{*}{IIoT} &  Industrial communication systems and their future challenges: Next-generation Ethernet, IIoT, and 5G.\\
\cline{1-1} \cline{2-2} \cline{4-4}

Nagpal et al. ~\cite{nagpal2019iiot} & 2019 & & IIoT based smart factory 4.0 over the cloud.\\
\cline{1-1} \cline{2-2} \cline{4-4}

Abuhasel et al. ~\cite{abuhasel2020secure} & 2020 & & A secure industrial Internet of Things (IIoT) framework for resource management in smart manufacturing.\\
\cline{1-1} \cline{2-2} \cline{4-4}

Nanda et al. ~\cite{nanda2020iiot} & 2020 & & IIOT based smart crop protection and irrigation system.\\
\cline{1-1} \cline{2-2} \cline{4-4}

Contreras et al. ~\cite{contreras2020implementing} & 2020 & & Implementing a novel use of multicriteria decision analysis to select IIoT platforms for smart manufacturing.\\
\cline{1-1} \cline{2-2} \cline{4-4}

Riasanow et al.~\cite{riasanow2021core}  & 2021 & & Examined the similarities of digital transformation in five platform ecosystems: automotive, Blockchain, financial, insurance, and IIoT. \\
\cline{1-1} \cline{2-2} \cline{4-4}

Jiang et al.~\cite{jiang20213gpp}  & 2021 & & 3GPP standardized 5G channel model for IIoT scenarios: A survey. \\
\cline{1-1} \cline{2-2} \cline{4-4}

Mantravadi et al.~\cite{mantravadi2022design}  & 2022 & & Design choices for next-generation IIoT-connected MES/MOM: An empirical study on smart factories.\\
\cline{1-1} \cline{2-2} \cline{4-4}

Fedullo et al.~\cite{fedullo2022comprehensive}  & 2022 & & A Comprehensive Review on Time Sensitive Networks with a Special Focus on Its Applicability to Industrial Smart and Distributed Measurement Systems. \\
\cline{1-1} \cline{2-2} \cline{4-4}

Lin et al.~\cite{lin2022novel}  & 2022 & & A Novel Architecture Combining Oracle with Decentralized Learning for IIoT. \\
\cline{1-1} \cline{2-2} \cline{3-3} \cline{4-4}

Harris et al.~\cite{harris2019decentralized} & 2019 & \multirow{13.5}{*}{\textcolor{black}{BC-AI}} & Decentralized and collaborative AI on Blockchain.\\
\cline{1-1} \cline{2-2}  \cline{4-4}

Wang et al. ~\cite{wang2019securing} & 2019 & & Securing data with Blockchain and AI.\\
\cline{1-1} \cline{2-2}  \cline{4-4}

Kumari et al. ~\cite{kumari2020blockchain} & 2020 & & AI-based approaches along with the advantages and challenges of integrating the BC technology and AI in the ECM system.\\
\cline{1-1} \cline{2-2}  \cline{4-4}

Ekramifard et al. ~\cite{ekramifard2020systematic} & 2020 & & A systematic literature review of integration of Blockchain and artificial intelligence.\\
\cline{1-1} \cline{2-2}  \cline{4-4}

Kumar et al. ~\cite{kumar2021integration} & 2021 & & An Integration of Blockchain and AI for secure data sharing and detection of CT images for the hospitals.\\
\cline{1-1} \cline{2-2}  \cline{4-4}

Alshamsi et al. ~\cite{alshamsi2021artificial} & 2021 & & Artificial intelligence and Blockchain for transparency in governance.\\
\cline{1-1} \cline{2-2}  \cline{4-4}

Jabarulla et al. ~\cite{jabarulla2021blockchain} & 2021 & & A Blockchain and artificial intelligence-based, patient-centric healthcare system for combating the COVID-19 pandemic.\\
\cline{1-1} \cline{2-2}  \cline{4-4}

Hua et al. ~\cite{hua2022applications} & 2022 & & Applications of Blockchain and artificial intelligence technologies for enabling prosumers in smart grids.\\
\cline{1-1} \cline{2-2}  \cline{4-4}

Aich et al. ~\cite{aich2022protecting} & 2022 & & Protecting personal healthcare record using Blockchain federated learning technologies.\\
\cline{1-1} \cline{2-2}  \cline{4-4}

Rahman et al. ~\cite{rahman2022blockchain} & 2022 & & Blockchain based AI-enabled Industry 4.0 CPS Protection against advanced persistent threat. \\
\cline{1-1} \cline{2-2} \cline{3-3} \cline{4-4}

Liu et al.~\cite{liu2019performance} & 2019 & \multirow{14.5}{*}{BC-IIoT} & Proposed a novel deep reinforcement learning (DRL)-based performance optimization framework for Blockchain-enabled IIoT systems.\\
\cline{1-1} \cline{2-2} \cline{4-4}

Gai et al.~\cite{gai2019differential} & 2019 & & Privacy-based Blockchain technology for internet-of-things in an industrial setting.\\
\cline{1-1} \cline{2-2} \cline{4-4}

Sengupta et al.~\cite{sengupta2020comprehensive} & 2020 & & A comprehensive survey on attacks, security issues and Blockchain solutions for IoT and IIoT.\\
\cline{1-1} \cline{2-2} \cline{4-4}

Wu et al.~\cite{wu2020convergence} & 2020 & & Convergence of Blockchain and edge computing for secure and scalable IIoT critical infrastructures in industry 4.0.\\
\cline{1-1} \cline{2-2} \cline{4-4}

Khattak et al.~\cite{khattak2020dynamic} & 2020 & & Application of Blockchain technology to introduce dynamic pricing in IIoT thereby enabling energy management in smart cities.\\
\cline{1-1} \cline{2-2} \cline{4-4}

Yu et al.~\cite{yu2021blockchain} & 2021 & & BC-enhanced data sharing with traceable and direct revocation in IIoT.\\
\cline{1-1} \cline{2-2} \cline{4-4}

Dwivedi et al.~\cite{dwivedi2021blockchain} & 2021 & & Blockchain-Based Internet of Things and Industrial IoT.\\
\cline{1-1} \cline{2-2} \cline{4-4}

Latif et al.~\cite{latif2021blockchain} & 2021 & & Presented a comprehensive survey on security issues, Blockchain architectures, and applications from the industrial perspective.\\
\cline{1-1} \cline{2-2} \cline{4-4} 

Mrabet et al.~\cite{mrabet2022secured} & 2022 & & A Secured Industrial Internet-of-Things Architecture Based on Blockchain Technology and Machine Learning for Sensor Access Control Systems in Smart Manufacturing.\\
\cline{1-1} \cline{2-2} \cline{4-4}

Huo et al.~\cite{huo2022comprehensive} & 2022 & & A Comprehensive Survey on Blockchain in Industrial Internet of Things: Motivations, Research Progresses, and Future Challenges.\\
\cline{1-1} \cline{2-2} \cline{3-3} \cline{4-4}

Sun et al.~\cite{sun2019ai} & 2019 & \multirow{12.5}{*}{AI-IIoT} & Introduced an intelligent computing architecture with cooperative edge and cloud computing for IIoT.\\
\cline{1-1} \cline{2-2} \cline{4-4}

Chen et al.~\cite{chen2019emerging} & 2019 & & Emerging trends of ml-based intelligent services for industrial internet of things.\\
\cline{1-1} \cline{2-2} \cline{4-4}

Miao et al.~\cite{miao2020blockchain} & 2020 & & Blockchain and AI-based natural gas industrial iot system: Architecture and design issues.\\
\cline{1-1} \cline{2-2} \cline{4-4}

Zheng et al.~\cite{zheng2020advancing} & 2020 & & Advancing from predictive maintenance to intelligent maintenance with AI and IIoT.\\
\cline{1-1} \cline{2-2} \cline{4-4}

Michailidis et al.~\cite{michailidis2020ai} & 2020 & & AI-inspired non-terrestrial networks for IIoT: Review on enabling technologies and applications.\\
\cline{1-1} \cline{2-2} \cline{4-4}

Rahman et al.~\cite{rahman2021ai} & 2021 & & AI-Enabled IIoT for live smart city event monitoring.\\
\cline{1-1} \cline{2-2} \cline{4-4}

Banaie et al.~\cite{banaie2021complementing} & 2021 & & Discussed the opportunities and challenges of realizing the AI, and particularly ML, approaches in IIoT systems that are used in the smart manufacturing environment.\\
\cline{1-1} \cline{2-2} \cline{4-4}

Ammar et al.~\cite{ammar2022implementing} & 2022 & & Implementing Industry 4.0 technologies in self-healing materials and digitally managing the quality of manufacturing.\\
\cline{1-1} \cline{2-2} \cline{4-4}

Raimundo et al.~\cite{raimundo2022cybersecurity} & 2022 & & Cybersecurity in the Internet of Things in Industrial Management.\\
\cline{1-1} \cline{2-2} \cline{4-4}

Ghayvat et al.~\cite{ghayvat2022strenuous} & 2022 & & STRENUOUS: Edge-Line Computing, AI and IIoT enabled GPS Spatiotemporal data-based Meta-transmission Healthcare Ecosystem for Virus Outbreaks Discovery.\\
\cline{1-1} \cline{2-2} \cline{3-3} \cline{4-4}

\end{tabular}
\end{table*}

\subsection{Related Surveys}
\textcolor{black}{The authors of \cite{hazra2021comprehensive} look into the issues and modern advancements of relevant state-of-the-art IIoT technologies, frameworks, and solutions for simplifying interoperability between distinct IIoT components. They also go over a number of IIoT standards, protocols, and models for digitizing the Industrial revolution that are compatible. However, they have identified certain obstacles and directions for their future work.
The authors of \cite{tange2020systematic} provide a systematic evaluation of the literature on IIoT Security from 2011 to 2019, with an emphasis on the IIoT's security requirements. Their secondary contribution is a discussion of how the relatively new Fog computing paradigm might be used to satisfy these needs and, as a result, improve IIoT security. This research \cite{kebande2022industrial} looks into the extent of IIoT forensics and the necessity of investigating and incorporating forensic standards and procedures into IIoT. The authors believe that the hypotheses presented here can serve as key building blocks for future IIoT-centered research frameworks.
Salih et al. \cite{salih2022comprehensive} focuses on the IoT's value in the industrial domain as a whole. It examines the IoT, focusing on its advantages and disadvantages, as well as some of the IoT's applications, such as in transportation and healthcare. In addition, the trends and statistics around IoT technology on the market are examined. Finally, the benefits of IoT technologies for COVID-19 are discussed, as well as the significance of IoT in telemedicine and healthcare. The authors are now working on an Industry 4.0 cybersecurity project, and the information in this paper is based on their findings. This study \cite{dhirani2021industrial, rahman2021study} presents a path for identifying, aligning, mapping, converging, and implementing the correct cybersecurity standards and techniques for safeguarding M2M communications in the IIoT, as well as an overview of best practices.
Atharvan et al. \cite{atharvan2022way} provide the state-of-the-art in networking layered frameworks for IIoT, evaluating links between cloud and edge computing technologies. They also looked into the kind of security attacks that could occur in an IIoT-driven 5G network, as well as how to prevent them. However, there are some flaws in their research. The authors of this paper \cite{jiang2021differential} perform a thorough examination of the prospects, uses, and limitations of differential privacy in the intelligent IIoT. They begin by reviewing related publications on IIoT and privacy protection. Then they look at industrial data privacy metrics and look at the conflict between using data for deep models and protecting individual privacy. A number of essential issues are summarized, and new study proposals are proposed. The authors of \cite{xenofontos2021consumer} presented a detailed analysis on the threats on the many stages of the IoT stack, such as device, infrastructure, communication, and service, together with the special properties of every layer that attackers can exploit.
They also use nine real-world cybersecurity incidents that targeted IoT devices in the consumer, commercial, and industrial sectors to highlight IoT-related vulnerabilities, exploitation techniques, attacks, impacts, and potential mitigation mechanisms and defense tactics.
By identifying and comparing distinct attacks on each IIoT layer, this research \cite{jayalaxmi2021taxonomy} assists researchers in understanding the reason for penetration. The major goal of this survey is to examine the numerous security concerns that IIoTs encounter and to give a comparative analysis of the current solutions to improve the protection systems of industrial IoT. This research also identifies several unsolved research issues for academia, technologists, and researchers in order to advance the IIoT area and its security features. The enhanced heuristics of the AOSR algorithm are presented in this work \cite{ud2021aosr}, along with specifics on how they improve performance efficiency in an SME-oriented warehouse. A detailed overview of the extensive validation using scenario-based experimentation and test cases explains how AOSR improved performance indicators by 60–148 percent in specific key warehouse regions.
The authors \cite{khalil2021deep} identified the critical ability of DL in the IIoT in this research. They begin by discussing several deep learning approaches, such as convolutional neural networks \cite{khan2021multinet}, auto-encoders, and recurrent neural networks, together with their applications in diverse environment \cite{khan2022accurate}. The authors further go over some DL application scenarios for IIoT systems, which includes intelligent manufacturing, metering, and agriculture. They outline a number of research problems associated to proper design and deployment of DL-IIoT. Eventually, they provided many potential future study directions in order to stimulate and inspire more study in this field.
This article \cite{soo2021recent} gives a thorough examination of IIoT security, including existing industry countermeasures, research suggestions, and ongoing difficulties. They classified IIoT technologies into the four-layer security architecture, assess implemented countermeasures based on CIA+ security criteria, report current countermeasure weaknesses, and identify open challenges and challenges. Moving to a data-centric approach that ensures data security whenever and wherever it travels could potentially solve industry deployment difficulties. Additionally, the summary of the related survey based on the considered technologies (BC, AI, and IIoT) and their integration has been cited in table \ref{tab:T1} together.}

\subsection{Contributions of the Survey}
\label{subsec:contributions}
{\color{black}
This survey discusses several leading technologies--BC, AI, and IIoT. In detail, we present an extensive analysis of their recent advanced concepts, security and confidentiality issues, tremendous applications, open issues, and further direction. We also adjust the integration's of these technologies aiming to great support applications in various areas. In this survey, we provide the subsequent contributions:

\begin{itemize}
      \item We present the state-of-art overviews and a general discussion regarding the BC, AI, and smart IIoT efficiently. 
    
      \item We discuss the integration benefits of these technologies-- BC-AI, and BC-AI-IIoT. We also searched their features and methods, which have already been established. 
    
     \item We highlight in-depth of security, stability, confidentiality, and reliability issues in the existing fields and upcoming fields, as well as using the BC-AI and IIoT technologies. Also, provided significant guidelines for the various solution tools and techniques against existing problems have been covered in this study. 
   
     \item We collaboratively and individually represent the applications of these technologies in various fields, including--BC-IIoT, AI, and AI-BC-IIoT applications.
    
     \item Finally, we cover the different issues, open challenges, opportunities and future guidance related to the considered technologies. 
\end{itemize}}

\begin{figure*}[h!]
    \centering
    \includegraphics[scale=1.3]{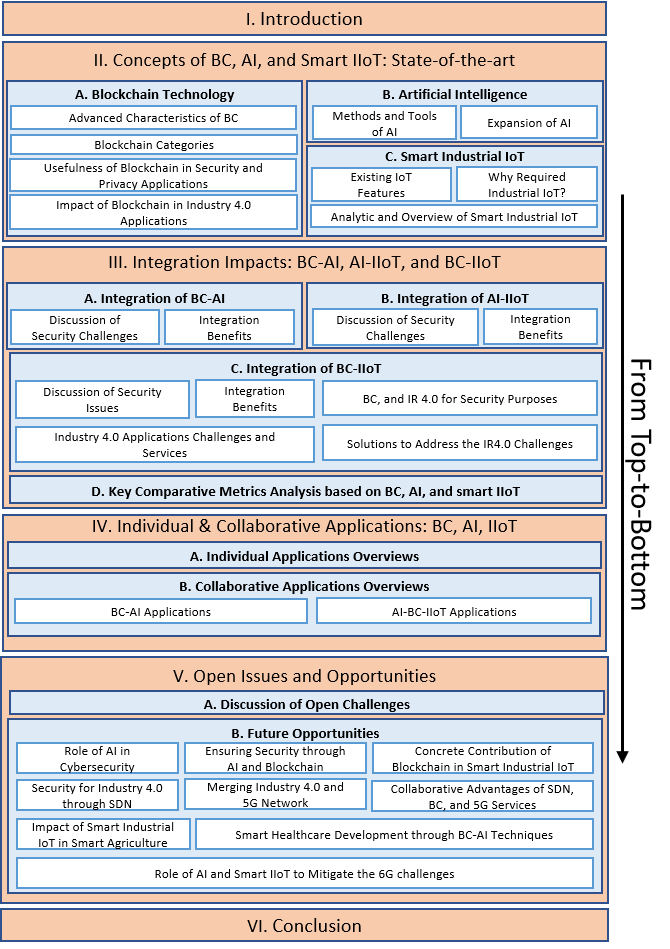}
   \caption{\textcolor{black}{Road map of this survey}}
    \label{fig:roadM}
\end{figure*}

\subsection{Outlines of this Survey}
\textcolor{black}{To the best of our knowledge, this study is the only potential and informative survey that incorporates all the considered technologies. Some of the notations are listed in Table \ref{tab:abbrtab}.
The remainder of this survey has been organized as follows: Section~\ref{sec:concepts} discuss the state-of-the-art overview regarding desired technologies--BC, AI, and smart IIoT individually.
Sections ~\ref{sec:taxo} describe the integration's impacts based on the BC-AI-IIoT technologies.
Further, the individual and collaborative applications have been covered in Section ~\ref{sec:tacol} on the basis of these technologies. After that, section~\ref{sec:opissfo} focuses on the research challenges and future opportunities of this survey. Lastly, we conclude this article in section ~\ref{sec:concl}}. For easy understanding, we add a complete road map of this survey, which is depicted in \ref{fig:roadM}.

\begin{table*}[h!]
\scriptsize
\caption{\textcolor{black}{State-of-the-art works focus on the utilization of BC, AI and IIoT in diverse techniques, major findings, and drawbacks \& challenges. The outcomes are reported in chronological order}}
\label{tab:T3}
\centering
\begin{tabular}{ |p{1.2cm}|p{2.5cm}|p{4.2cm}|p{3.8cm}|p{2.5cm}|}
\hline
\textbf{Works}  & \textbf{Methods and Tools} & \textbf{Major Contributions} & \textbf{Drawbacks and Challenges} & \textbf{Applications Fields} \\
\hline

Khan et al. ~\cite{khan2021federated} (2022) &  Wireless networking, FL, IoT &  Identification of challenges and advantages of FL in healthcare predictive modelling and addressing critical issues related to privacy and ownership are addressed & The study requires further analysis with hyper-parameters. & Education, Healthcare, and Industries\\
\hline

Rani et al. ~\cite{rani2022blockchain} (2022) & Neural Network (SqueezeNet, ResNet50, VGG19, InceptionV3), Transfer learning & To develop a secured protocol for communication by integration Blockchain and IoT for healthcare system monitoring & \textcolor{black}{Data needs to be transmitted that still may include some threats to be hacked.} & Healthcare\\
\hline

Verma et al. ~\cite{verma2022security} (2022) & Blockchain, IoT &  Reviewed security threat to the IoT based systems and presented Blockchain-based mitigation process & The authors presented Blockchain based security enhancement system, however, Blockchain based security system can also be under attack. & IoT security \\
\hline

Huo et al. \cite{huo2022comprehensive} (2022) & Blockchain, IIoT & Scope and challenges of integrating Blockchain in IIoT sector & \textcolor{black}{This article lacks taxonomy of related research works.} & IIoT manufacturing challenges\\
\hline

Rajawat et al. \cite{rajawat2022ai} (2022) & AI and Blockchain for data security & Blockchain and AI-based model for enhancing the data security in the smart applications & \textcolor{black}{Proper adaptation of healthcare is missing.} & Smart healthcare and Smart City\\
\hline

Chatterjee et al. \cite{chatterjee2022security} (2022) & IoT, Blockchain & Data security issues in IoT based communication system & \textcolor{black}{Addressed major security issues in the IoT based communication system with possible solution.} & IoT security\\
\hline

Ullah et al. \cite{ullah2021ai} (2021) & IIoT, VD-Net, AI  & Industrial Surveillance system with IIoT and VD-Net & \textcolor{black}{Considered available dataset but with different view of same images this method has not been tested.} & Industrial Security (surveillance) \\
\hline

Zhang et al. \cite{zhang2021bc}(2021) & IIoT, Blockchain, Federated Learning & A secured data transmission system by integrating Blockchain and Federated learning model in IIoR scenario & \textcolor{black}{The cost of complexity is not examined to a significant extent.} & Security of data transmission\\
\hline

Xiao et al. \cite{xiao2021collaborative} (2021) & IIoT, DNN, Cloud & Integration of AI and DNN with a cloud edge-based system that takes into account both energy consumption and quality of service  &\textcolor{black}{Cost analysis is missing.} & Computing service \\
\hline

Ramasamy et al. \cite{ramasamy2021data} (2021) & AI, Blockchain, IoT advancement & Focuses on the integration of AI, Blockchain, and IoT with independent plans of action  &\textcolor{black}{Only highlighted the data security.} & Safe sharing of information \\
\hline

Mehta et al. \cite{mehta2021blockchain} (2021) & Blockchain & Secured contract transaction scheme in IIoT scenario & \textcolor{black}{Considered only industrial perspective and in real scenario transaction also occurs with other pirates such as landowners or service providers.} & Supply chain\\
\hline

Espinoza et al. \cite{espinoza2021design} (2021) & AI and smart IIoT & Excellent combination between AI and IIoT for vibration-based fault diagnosis architecture & \textcolor{black}{CVCM-based optimized algorithm should be offered for the data collection.} & Global positioning system location and Industry 4.0 applications\\
\hline

Zhao et al. \cite{zhao2021secure} (2021) & Edge Computing, AI, and Industrial IoT for security purposes & An edge computing framework for intelligent IIoT AI-driven to improve the flexibility and security of IIoT edge systems & Resource utilization approaches need to be clarified more concisely. & Real-time performance analysis for smart IIoT \\
\hline

Rahman et al. \cite{rahman2020distb} (2020) & Blockchain. IR 4.0 & Blockchain for securing data and enhancing privacy & \textcolor{black}{As Data is transferred, there still may remain some security threats, handling those attacks not mentioned.} & Security for IIoT based on Blockchain and SDN \\
\hline

Singh et al.~\cite{singh2020convergence} (2020) & Blockchain, AI, and IoT Intelligence & Discussed the sustainability of smart city with the support of AI and Blockchain technology & \textcolor{black}{Development of an intelligent community should have been discussed.} & Smart city\\
\hline

Wu et al. \cite{wu2020convergence} (2020) & Blockchain, Edge computing, IIoT & Integration of Blockchain and Edge Computing for enhancing security of IIoT based systems & \textcolor{black}{Comparison with other related research not presented.} & Security and privacy for smart IIoT.\\
\hline

Guan et al. \cite{guan2020towards} (2020) & Blockchain, IIoT & Blockchain-based secured energy trade system & \textcolor{black}{When data is transmitted there still may some security threats in the overall system that is not considered in this work. }& Energy trading \\
\hline

Sharma et al. \cite{sharma2020role} (2020) & Blockchain, AI, and IoT & Presented a Smart Road Traffic Management System (SRTMS) architecture for reducing vehicle collision with the help of emerging technologies AI, Blockchain, and IoT & \textcolor{black}{Collision detection and correction algorithm is messing. } & Road traffic management \\
\hline

Wazid et al. \cite{wazid2020private} (2020) & Drone technology, AI, Blockchain & AI and Blockchain based secured systems for healthcare professionals &\textcolor{black}{Healthcare based data are susceptible and private. Collecting data and additionally providing security to them is challenging.} & Healthcare \\
\hline

Campero et al. \cite{campero2020smart} (2020) & AI (SVM, CNN, Naive Bayes), smart IIoT & Design of a smart Helmet that can sense the surroundings, monitor, and safeguard employees in any dangerous circumstance & \textcolor{black}{Maintaining the data transmission quick and smooth is challenging.} & Industrial construction sites\\
\hline
\end{tabular}
\end{table*}

\section{Concepts of BC, AI, and Smart IIoT: State-of-the-art} \label{sec:concepts}

\subsection{Blockchain Technology}
\subsubsection{Advanced Characteristics of BC}
\textcolor{black}{Asymmetric cryptography \cite{olwenyicryptography} is a strategy for dealing with some Blockchain concepts connected to AI and IoT. Public and private keys are a fundamental principle of asymmetric cryptography. We begin by encrypting a numerical message M. Max, for example, wishes to deliver M to Mary devoid of anyone being capable of accessing it. Public and private keys of Max are indicated $\left(K_{p w}^{M a x}, K_{p r}^{M a x}\right)$, and Mary's ones $\left(K_{p u}^{M a r y}, K_{p r}^{M a r y}\right)$. Max has the ability to cipher $M$ with the use of an encryption function $C$ and public key of Mary's:}
\begin{equation}
V=C\left(K_{p u}^{\text {Mary }}, M\right)
\end{equation}
\textcolor{black}{Mary can interpret the information with the use of her private key and the decoding function D associated with C once she has received the value V:}
\begin{equation}
M=D\left(K_{p r}^{\text {Mary }}, V\right)
\end{equation}
\textcolor{black}{The identification of a message M is the second question: The ability of Mary to detect that Max is the one delivering the message. The following is the solution: Max dispatches both M and the signature S to Mary where:}
\begin{equation}
S=C\left(K_{\rho r}^{\operatorname{Max}}, M\right)
\end{equation}
\textcolor{black}{After Mary has received $(S, M)$, she has the ability to decode $S$ with Max's public key:}
\begin{equation}
m=D\left(K_{p u}^{M a x}, S\right)
\end{equation}
\textcolor{black}{The most important distinction is the fact that the private key of Max should be kept private at all times: only the authorized person of the key should have access to it, but the public key may be delivered to anyone. The RSA algorithm is the most well-known asymmetric encryption technique: it uses prime numbers to create public and private keys and generate mathematical functions (power functions and modules) to encipher and decrypt messages.}

\vspace{2mm}
\subsubsection{Blockchain Categories}
\begin{figure}[h]
    \centering
    \includegraphics[scale=0.22]{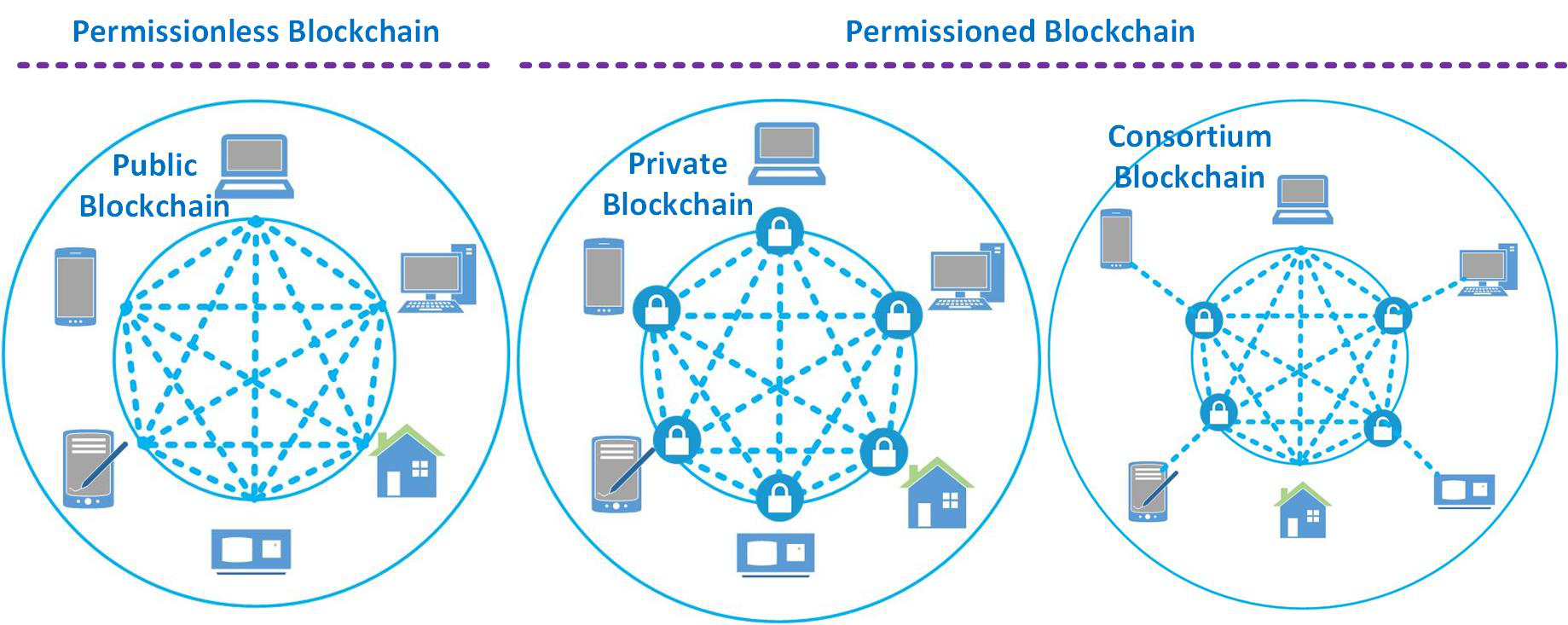}
   \caption{\textcolor{black}{Types of Blockchain}}
    \label{fig:f1}
\end{figure}

In accordance with user accessibility, a Blockchain can be categorized into two types that shown in Fig. \ref{fig:f1}.
Open Blockchain, also known as Public Blockchain, permits interaction between users with deceptive identities. Sometimes, the identity of the users involved in an agreement remains completely anonymous. Public Blockchain technology imposes very little in the way of privacy. 
In terms of data accessibility, private Blockchain technology provides a greater level of reliability and privacy. 
Another accessible BC is consortium BC. This particular type of Blockchain has authorized nodes to manage decentralized ledgers. 

\vspace{2mm}
\subsubsection{Usefulness of Blockchain in Security and Privacy Applications}
Blockchain can be used to validate the user in the network system of the 5th-century \cite{unal2020policy}. The data alteration of attackers becomes difficult as Blockchain stores the information along a couple of computers together that creates a network. Additionally, the smart contracts in the Blockchain mechanism enable the establishment of authentic and verified communication along the 5G network \cite{nguyen2020blockchain}. 
For the sensors used in IoT, Blockchain can be an option to provide an encryption mechanism and an environment for trustworthy execution and identification \cite{rathee2021secure}. 
\textcolor{black}{The essential steps in the BC-based distributed AC model for IoT are as follows:\newline
Generate (G1, G2, e) with k as input parameters, where G1 and G2 are the order groups of e and q, respectively, and satisfy G1 × G2→G2; randomly pick the master key $\mathrm{s} \in \mathbb{Z}_{\mathrm{q}}^{*}$, which is one of the system's two secure hash functions \cite{saleem2020pata, saleem2021fuzzy}.}
\begin{equation}
\begin{aligned}
&H A S H_{1}=\{0,1\}^{*} \rightarrow G_{1}^{*} \\
&H A S H_{2}=G_{2} \rightarrow\{0,1\}^{n}
\end{aligned}
\end{equation}
 \textcolor{black}{Using a self-generated address and the device's identity flag ID, each IoT device in the system can request an attribute value pair (prop ni, prop vi) from the AAA. The IoT device first chooses $ \mathrm{s} \in \mathbb{Z}_{\mathrm{q}}^{*} $ as the device's key SKi, then the public key PKi = kG that corresponds to this key. The IoT device can hash the PKiIDdevtimestamp to acquire the Blockchain address corresponding to this public key, and then encode the BC address obtained using the Base58Check function, as follows:} 
\begin{equation}
\text { Address }=\text { Base } 58 \mathrm{Check}\left(\mathrm{H}_{2}\left(P K_{i}\left\|D_{\text {dev }}\right\| \text { timestamp }\right)\right)
\end{equation} 
This address is used to request an attribute-value pair for an IoT device from an attribute authorization authority.
\textcolor{black}{When an IoT device requests an attribute value pair (prop ni, prop vi), the appropriate attribute authorization authority must verify on a case-by-case basis whether the IoT device is allowed to have the attribute value pair. If this verification is successful, this attribute authority generates an authorization transaction for the attribute value pair right away, concatenates the transaction with the current timestamp, performs a hash operation to obtain a hash value, and then signs the hash value as follows:}
\begin{equation}
\operatorname{Sig}_{S K_{\mathbf{i}}}\left(H_{1}\left(A A \stackrel{\left(\text { prop }_{n_{i}}, \text { prop }_{v_{i}}\right)}{\longrightarrow} \text { Address } \| \text { timestamp }\right)\right)
\end{equation}
Finally, the attribute authorization authority bundles the signature value, transaction, and current timestamp into a single package and uploads it to the federated Blockchain.
\textcolor{black}{When an IoT device is registered with an attribute authorization authority, the attribute authorization authority must create a reasonable access control policy file based on the specific situation, after which the transaction and the current timestamp are concatenated and hashed, and the hash value is signed, as follows:}
\begin{equation}
\operatorname{Sig}_{S K_{I D}}\left(H_{1}\left(A A \stackrel{P_{I D}}{\longrightarrow} \text { Address }|| \text { timestamp }\right)\right)
\end{equation}
Moreover, the AAA will use Fabric to combine the signature value, transaction, and current timestamp.
\textcolor{black}{Because the IoT system is dynamic and evolving, access control policies for IoT terminals may fail due to changes in business requirements. In this case, the AC policy of the IoT device must be revoked by connecting the transaction with the current timestamp for the operation and hashing it, then signing the hash value, as follows:}
\begin{equation}
\operatorname{Sig}_{S K_{I D}}\left(H_{1}\left(A A \stackrel{P_{I D}}{\longleftarrow} \text { Address } \| \text { timestamp }\right)\right)
\end{equation}
The attribute authorizer will bundle the signature value, transaction, and current timestamp into a single package and send it to the federated Blockchain.
Lastly, because of the cryptographic feature of the technology, security and privacy are preserved in Blockchain technology.

\subsubsection{\textcolor{black}{Impact of Blockchain in Industry 4.0 Applications}}
Blockchain can be a possible and suitable solution for the 5G networking scheme for IR 4.0. With the encryption properties of Blockchain, user validation is possible at both ends between parties of the IR ecosystem. Again for the decentralization of the network, data is stored in the decentralized server with better authorization and authentication mechanism, which is indeed a robust method for storing data along the cloud \cite{serrano2021blockchain}. 

\subsection{Artificial Intelligence}
\begin{figure}[h]
    \centering
    \includegraphics[scale=0.29]{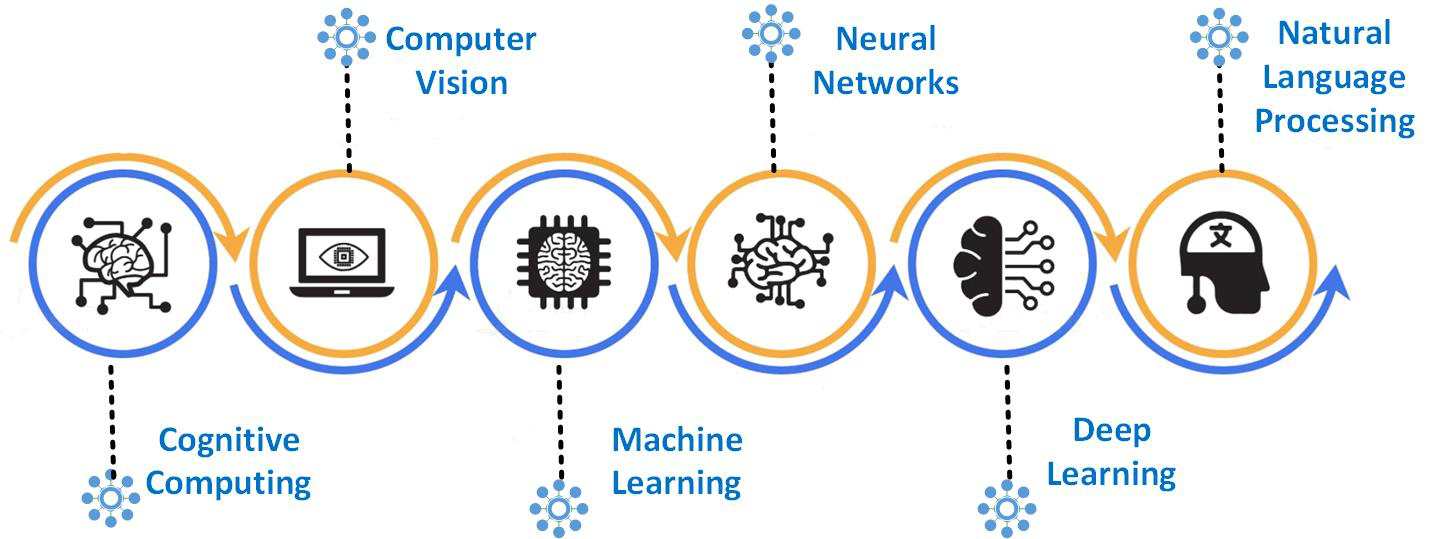}
   \caption{Feature of Artificial Intelligence}
    \label{fig:f2n}
\end{figure}

Recent developments in the world of computative resources and programming coupled with the significant growth of available data on the web means that there is an extensive variety of applications of artificial intelligence in several sectors of our everyday life. 
A simple scenario of AI is shown in Fig. \ref{fig:f2n}.
Some of the key features of AI includes--
 Deep Learning is one of the primary characteristics of AI, an ML algorithm that trains computers to perform tasks by studying human examples. 
 Facial recognition is another prominent feature of AI through different model or networks \cite{Debnath2022}. 
 AI also has the property of performing tasks in an automated manner. 
Other characteristics of AI are chat-bots which are essential software used for answering client's issues via text. 
Artificial intelligence also possesses the ability to assist in performing problems related to quantum physics with increased accuracy. It performs these complex problems with the assistance of a neural network. This has enabled a revolution in the development made in this particular field.
Finally, a key characteristic of AI is cloud computing. The increased amount of digital data from a wide range of sources requires a storage facility.


\vspace{2mm}
\subsubsection{Methods and Tools of AI}
\textcolor{black}{AI has the ability to transform manufacturing into smart industries based on data-driven decision-making. AI is being implemented to mitigate the risks of machine failure, increase quality control and productivity, and achieve economic means of manufacturing 
Several types of research explored the potential of AI  in smart industries. 
Recent developments in using AI in smart manufacturing are also presented by authors in \cite{preuveneers2017intelligent}. 
A detailed summary of AI  and big data in the field of Industry 4.0 is reviewed in \cite{jagatheesaperumal2021duo}. Artificial Intelligence is seen as a significant driver of the Industry 4.0 revolution, as identified in \cite{9952633}. The research introduces a novel way of generating a map of Industry 4.0 technologies with the help of natural language processing to gather technical information from relevant research articles. Various areas like computer science, mathematical analysis, statistical data, biological science, human sociology, and many other areas of science have participated to consolidate AI's multidisciplinary attributes. The overall data created from each of these domains instigates towards knowledge. It is critical to analyze this data in order to uncover the concepts that underpin it. It is simple for the human brain to perform, but it requires a longer time period. The reason behind this is due to the fact that information in reality has some unsuitable features: (i) massive volume, (ii) indeterminate nature, (iii) diverse information sources, (iv) requires simultaneous processing, (v) varies constantly. Further attributes comprise unreliability, spontaneity, and so forth. AI may be identified as a method for effectively using the information in a manner that is comprehensible to those who supply it, adjustable (when an error occurs), beneficial in the present situation, and reasonable. Because of this, data science approaches are heavily used in AI. Here, data science is the science of generating elements and techniques for studying significant quantities of data and deriving information from them. In summary, the various methods and tools of Artificial Intelligence are Machine Learning, ANN, DT, Natural Language Processing, Automation and Robotics, and SVM.} 

\vspace{2mm}
\subsubsection{Expansion of AI}
\textcolor{black}{The growth of AI in response to the ever-altering scenario is a significant source of concern for everyone since it raises a variety of challenges ranging from administrative initiatives to private usage. On top of assisting to make our daily life more easier, AI also has several uses.
 In order to boost economic growth and productivity and maintain a moderate and steady market, AI is used. AI assists businesses in expanding. Furthermore, AI assists employees in the workplace by providing convenience.
The programs deployed by AI to do the assigned work are far more effective than those used by employees and the task will be completed significantly faster. It would provide far more consistent findings when evaluating algorithm efficiency.
AI further focuses on the algorithms' clarity and liability. For example, one must safeguard highly skilled data such as the patient's medical report without jeopardizing data sensitivity.
}

\subsection{Smart Industrial IoT}
\subsubsection{Existing IoT Features}

One of the most significant features of IoT remains communication which is critical for establishing a secured connection between sensors (and other relevant devices) to the IoT platform via the server. The established communication is bidirectional, and high-speed communication is required for exchanging information amongst the several interconnected devices and the server. In order to effectively establish a communication medium and to help understand the IoT platform regarding the source of the data, it is essential to identify and manage the device endpoint. IoT also offers an analysis feature which assists to study the collected data and make intelligent or smart decisions. The IoT platform performs real-time analysis in regard to data aggregation, filtering of data, and successive correlation. Integration is yet another feature of IoT which enables the system to improve its performance. IoT further permits the connectivity between heterogeneous devices and networks. The devices in the IoT platform are able to communicate amongst each other despite their heterogeneity, thereby establishing a smart system. Finally, the IoT platform also performs the action based on data collection, communication, and intelligent decision.

\subsubsection{Why Required Industrial IoT?}
\begin{figure}[h]
    \centering
    \includegraphics[scale=0.29]{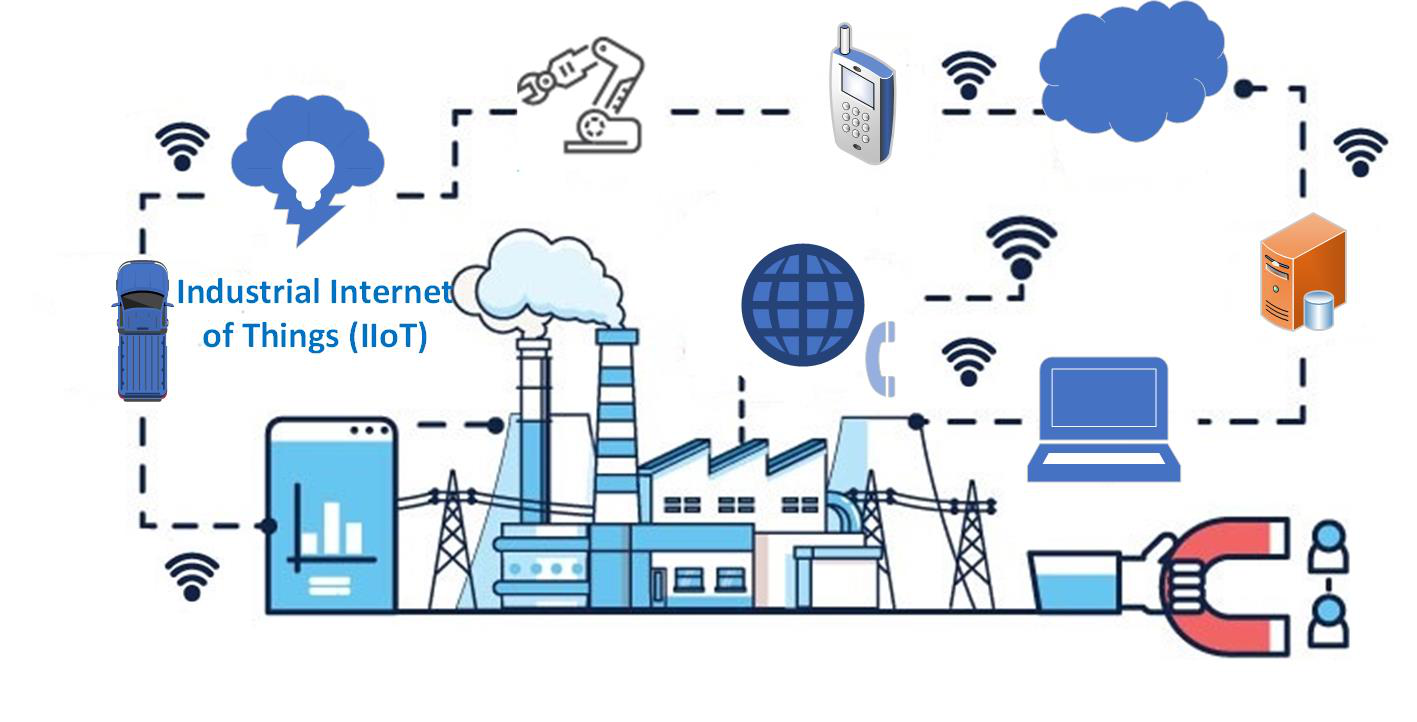}
   \caption{Smart Industrial IoT Architecture}
    \label{fig:f3n}
\end{figure}

The IIoT platform involves interconnection between smart devices through a distributed network. These smart devices range from sensors, automated technologies, and relevant devices to highly efficient gateways coupled with simultaneous data analysis. Fig. \ref{fig:f3n} has been deemed to demonstrate smart IIoT for a better understanding of readers. However, IIoT has the potential towards a transformative solution throughout several sectors including smart cities, smart healthcare, robotics and automation, intelligent transport system and many more.

\subsubsection{Analytic and Overview of Smart Industrial IoT}
IoT technology plays a significant function in advanced industries by automating tasks and eventually reducing relevant expenses. IIoT has gained increased attention from both the industrial and academic sectors as state-of-the-art technology for the process of manufacturing in an industrial environment. IIoT simply refers to the interconnection between several sensors, tools, and relevant devices, which are merged with computer applications to perform surveillance, development, and energy supervision. IoT features increased control of operation by providing machine operators with up-to-date information and increased reliability. 

\textcolor{black}{In one example, the SMART strategy is a method for dealing with multi-criteria decision making (MCDM) challenges used in this paper \cite{alghofaili2022trust}. It depends upon the idea that every other possibility is made up of several standards with principles, and every benchmark has a weight that reflects its importance in relation to other benchmarks \cite{oktavianti2019simple}. A combination of the SMART and Shannon's entropy approaches is utilized in some studies to determine weights depending on the provided criteria. The following three equations can be used to compute Shannon's entropy (Ej). Assume $kj (j=1,2,3 \ldots)$ contains numerous possibilities and ki (i= $1,2,3 \ldots)$ indicates the criteria contained in these alternatives. Kij then indicates the ith benchmark value in the jth alternative, forming the weight evaluation mechanism on this premise. These elements should be standardized using:}
\begin{equation}
R i j=\frac{k_{ij}}{\sum_{i=1}^{m} \sum_{i=1}^{n} k_{ij}}
\end{equation}
\textcolor{black}{where Rij is the specific gravity per kilogram and $m$ is the number of criteria. Then, the entropy for each factor alternative Ej is then calculated as follows:}
\begin{equation}
E_{j}=\left[\frac{-1}{\ln (m)}\right] \sum_{i=1}^{m}\left[R_{ij} \ln \left(R_{i j}\right)\right]
\end{equation}
\textcolor{black}{where m is the matrix's number of standardized assessment possibilities and ij is the number of criteria.}
\begin{equation}
D_{j}=1-E_{j}
\end{equation}
\textcolor{black}{where, Dj denotes the diversity criterion.\newline
Using the mean absolute error (MAE) to calculate the dynamic threshold (DT). MAE is an element that quantifies the accuracy of forecasts as compared to real results. MAE is used as it provides a simple method for measuring the content of errors \cite{reich2016case}. It is frequently utilized in the privacy area to compute errors based upon the issue.}
\begin{equation}
\text { Dynamic Threshold }(\mathrm{DT})=\frac{\sum_{i=1}^{n}\left|u\left(a_{i}\right)-\overline{u\left(a_{i}\right)}\right|}{n}
\end{equation}
\textcolor{black}{where $u\left(a_{i}\right)$ is the value of the trust, $\overline{u\left(a_{i}\right)}$ is the predicted trust value, and $n$ is the total number of samples.\newline
The comparison of the value of trust and the DT calculated generates a value and if the value of trust is higher or equal to the DT value, the device is trusted; or else, it is considered untrusted:}
\begin{equation}
\text { Trust Score }=\left\{\begin{array}{l}
u\left(a_{i}\right)<\mathrm{DT}, \text { Untrused } \\
u\left(a_{i}\right) \geq \mathrm{DT}, \text { Trusted }
\end{array}\right.
\end{equation}

To summarise this section, we consider the state-of-the-art table \ref{tab:T3}; this table provides a piece of tremendous information based on the BC, AI, and intelligent IIoT technologies. Also, this table offers some parameters such as--tools and methods, main contribution, limitations, and application areas as well based on the existing research.

\begin{table*}[h!]
\scriptsize
\caption{\textcolor{black}{Previous Studies Analysis based on Interrogation and Motivation of BC-AI-IIoT technologies}}
 \label{tab:T4}
\centering
\begin{tabular}{ |p{2cm}|p{3.0cm} |p{6cm}|p{4cm}|}
\hline
\textbf{Authors \& Year} & \textbf{Integrated Techniques} & \textbf{Main Motivations} & \textbf{Limitations} \vspace{1mm} \\ 
\hline
	
Karamchandani et al. \cite{karamchandani2022integrative} (2022) & AI, Robotics, IR 4.0 applications, Blockchain & In the context of Big Data, integration benefits of BC. & \textcolor{black}{Security issues of Blockchain in the Big data scenario.} \\
\hline

Wolf et al. \cite{wolf2022potential} (2022) & AI & GHG emission reduction in the healthcare based scenarion by integrating AI based methods. & \textcolor{black}{Performance matrices not discussed }\\
\hline

Deebak et al. \cite{deebak2022lightweight} (2022) & IoT, Blockchain, AI & Blockchain integrated secured user authentication system. & \textcolor{black}{In real time the model is not tested.}\\
\hline

Kumar et al. \cite{kumar2022artificial} (2022) & Blockchain, AI, IIoT & The effects of Blockchain and AI in the sector of Business. & \textcolor{black}{considered theoretical context only.}\\
\hline

Panagopoulos et al. \cite{panagopoulos2022incentivizing} (2022) & AI, IoT & Data transmission flow management with AI based methods. & \textcolor{black}{Proposed method need to be tested upon some real time healthcare data}\\
\hline

Sarosh et al. \cite{sarosh2022efficient} (2022) & AI, Image Encryption & Image encryption method for secured transmission through AI based IoT system. & \textcolor{black}{Security issues of AI based image encryption mecahnism}\\
\hline

Alahmari et al. \cite{alahmari2022musawah} (2022) & AI, ML, NLP, Big data & A model to identify healthcare services by analyzing social media. & \textcolor{black}{Attack scenario not considered in this work}\\
\hline

\textcolor{red}{}Khan et al. ~\cite{khan2021federated} (2021) &	AI, IoT, NFV, 5G & A study the application of SDN-NFV for the advancement of the present world and additionally discussed some issues related to this. & \textcolor{red}{}Considered only cases related to rational agents and attacks that are already settled only.\\
\hline

Kumar et al. \cite{kumar2021integration} (2021) & Blockchain, Deep learning & Combination of deep learning with Blockchain for CT image for healthcare scenario. & \textcolor{black}{Dataset consists 5842 CT cases only.}\\
\hline

Alrubei et al. \cite{alrubei2021use} (2021) & AI, IoT, Blockchain & Data security enhancement by integrating Blockchain with IoT. & \textcolor{black}{System with real time response time need further analysis.} \\
\hline

 Yu et al. \cite{yu2021blockchain} (2021) & Blockchain, IIoT &  To validate identity in a system based on the combination of Blockchain and IIoT. & \textcolor{black}{The transaction size considered in this work is not very high.}\\
 \hline

Islam et al. \cite{islam2021blockchain} (2021) & Blockchain, IoT, NFV, SDN & Cluster head selection algorithm for IoT based smart city energy management. & \textcolor{black}{Analysis of cyber attacks requires further analysis with the mitigation procedure.}\\
\hline

Luo et al. \cite{luo2021great} (2021) & Blockchain, Deep neural network & Integration of Blockchain and AI to enhance the performance of business operations. & \textcolor{black}{Complexity analysis of implementing such technology requires further analysis.}\\
\hline

Singh et al. \cite{singh2020blockiotintelligence} (2020) & Blockchain, IoT & An effective and intelligent way of integrating Blockchain with AI-IoT based scenario. & \textcolor{black}{Lacks of comparison with other related research.}\\
\hline

Hu et al. \cite{hu2020securing} (2020) & SDN, Blockchain, IoT & Rewarding system for game theory by integrating Blockchain technology. & \textcolor{black}{Attack in such scenario not examined vastly.}\\
\hline

Yazdinejad \cite{yazdinejad2020energy} (2020) & Blockchain, SDN, IoT & Energy efffcient method for transferring data witin the IoT devices. & \textcolor{black}{With real time datasets the management of data needs further analysis.}\\
\hline

Wang et al. \cite{wang2021optimized} (2021) & Blockchain, IIoT & Verification of transaction process securely with Blockchain. & \textcolor{black}{Blocks can store 2000-5000 transaction that may lead to 51\% attack or some other security attacks.}\\
\hline 

Ghahramani et al.  \cite{ghahramani2020ai} (2020) & AI, CPS, NN, IIoT, Genetic Algorithm & Effective manufacturing process by investing various perspectives. & \textcolor{black}{Lacks comparison with other research methods.}\\
\hline
Rahman et al. \cite{rahman2020block} (2020) & Blockchain, SDN, IoT, OpenFlow & Enhancement of security by integrating Blockchain with SDN. & \textcolor{black}{Latency and complexity not analyzed in this work.}\\
\hline

Liu et al. \cite{liu2019performance} (2019) & Blockchain, IIoT  Reinforcement learning & Development of a unique deep reinforcement learning (DRL)-based performance optimization methodology for Blockchain-enabled IIoT systems. & \textcolor{black}{Lacks performance comparisons with other studies.}\\
\hline
\end{tabular}
\end{table*}

\section{Integration Impacts: BC-AI, AI-IIoT, and BC-IIoT}\label{sec:taxo}

\subsection{Integration of BC-AI}

\subsubsection{Discussion of Security Challenges}
\textcolor{black}{Security concerns are seen as one of the most important issues to address in smart apps, particularly in the case of Blockchain-based applications, issues such as security or preserving users' private information are a top priority. }
\begin{itemize}
    \item \textcolor{black}{With the increasing creation of data from many sources, data security has become a major security problem for many organizations. They have become a target for attackers due to the large amount of data they generate. Again, Blockchain technology alone will not be able to uncover the flaws because the review process would be lengthy.}
    
     \item  \textcolor{black}{Through the core mechanism of Blockchain, which is encryption and hashing, the adoption of Blockchain technology in the field of data transmission and generation-based systems unquestionably strengthens security. In the Blockchain system, the account address is utilized to perform transactions. As a result, there is a certain amount of obscurity.}
     
     \item \textcolor{black}{The majority of devices in IR 4.0 based systems with IoT integration have limited resources. With this limited resources, the implementation of Blockchain is a major challenge as it contains vast amount of parameters and blocks of data.}
     
     \item \textcolor{black}{Another challenge with combining BC with AI is the scalability of Blockchain. For example, bitcoin transactions are deemed low since the typical block time is 10 minutes and only 5 to 7 transactions occur per second. }
     
     \item \textcolor{black}{Private Blockchains have a quicker transaction rate than public Blockchains, but they do not provide totally decentralized networks. To provide Byzantine fault tolerance, the private Blockchain consensus algorithms contain a round of voting. This shouldn't be used on public Blockchain networks. The public Blockchain philosophy assumes that all users are equal and that there is no central authority.}

\end{itemize}

\subsubsection{Integration Benefits}
\begin{figure}[h]
    \centering
    \includegraphics[scale=0.27]{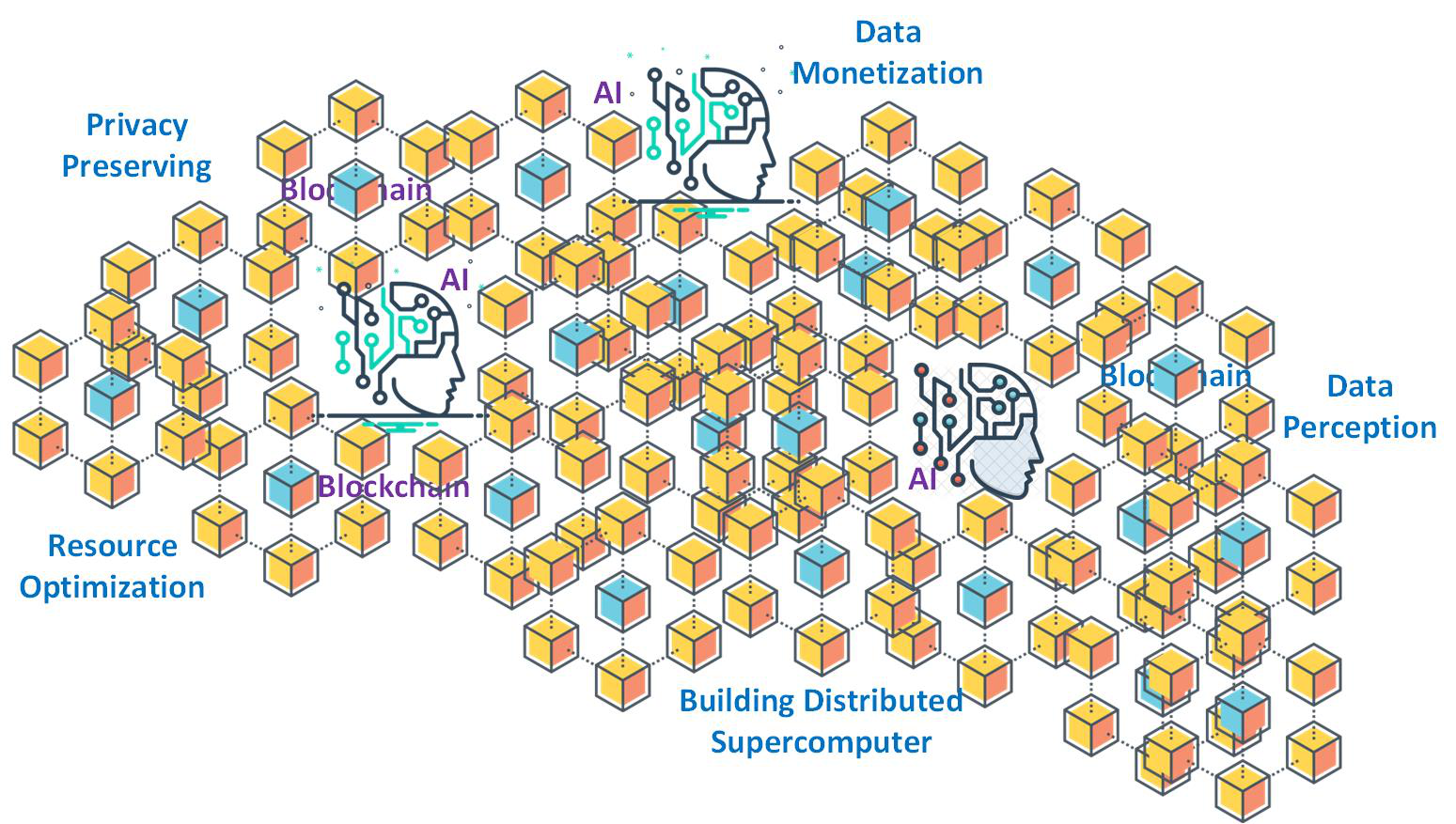}
   \caption{\textcolor{black}{Integration of BC-AI Approaches}}
    \label{fig:f3}
\end{figure}

\textcolor{black}{Integrating Blockchain and artificial intelligence strengthens and empowers each other. A combined approach has been depicted in Fig.\ref{fig:f3}. There are two ways to define it. Blockchain assists AI become more efficient and robust, and likewise.}

\textbf{Blockchain for AI:}

\begin{itemize}
    
    \item \textcolor{black}{\textbf{Data source transparency:} Integrating several technologies is primarily intended to provide a secure environment for data sharing. There are a variety of approaches to guaranteeing data transparency. Blockchain has acquired popularity as a result of its consensus process for guaranteeing data security while still retaining data openness. The data to be transferred is added to a chain of blocks where the nodes are synchronized to enable data traceability during the transaction using this approach. As a result, Blockchain technology may be used to create a transparent data exchange medium.}

    \item \textcolor{black}{\textbf{Fair Rewarding Mechanism:} Proper provider monitoring is available because to Blockchain technology. In this situation, the provider that does not follow the system's rules in the case of a transaction is penalized, while honest suppliers are rewarded \cite{tang2019incentivizing}.}

    \item\textcolor{black}{\textbf{Decentralization:} The idea of a central server is replaced with a distributed ledger in the Blockchain system. The authority of a single central server is obliterated, and the system's management power is no longer under the control of a single entity.}

    \item\textcolor{black}{\textbf{User's Data Confidentiality:} If not adequately protected, the large volume of data shared in today's online environment may result in data leakage. User privacy may be ensured through the Blockchain using traditional mechanisms such as pseudonyms or shuffling. These approaches, however, cannot guarantee complete secrecy because intruders can use any mining process to obtain the user's personal information. Incorporating conditional anonymity with the ring signature can strengthen the confidentiality measure and assure sufficient privacy \cite{zhang2022novel, 9333523}. }
    
    \item \textcolor{black}{\textbf{Power distribution:} The AI based mechanism provides computing power by single unit. However, with the integration of IR 4.0 technologies, the amount of data sharing and receiving has been rapidly increased, increasing the complexity of computation. Blockchain decentralization mechanism can mitigate this issue through the distribution property. }
\end{itemize}

\textbf{AI for Blockchain:}
\textcolor{black}{The Blockchain-based system is complicated since it involves a large number of security methods, settings, and decentralization processes. As a result, AI-based approaches may aid in the system's organization in a simpler manner, allowing for the efficient use of the Blockchain method while achieving improved accuracy and a reduction in mistakes caused by man-made impacts.}

\begin{itemize}
    \item \textcolor{black}{\textbf{Reliability:} For boosting security, Blockchain-based approaches are highly popular. The organization suffered a significant loss as a result of this \cite{alsirhani2022securing}. To maintain a safe environment, AI might be a solution that generates automatic contracts for determining smart contract flaws. Thus, by combining AI with Blockchain, a durable and safe system may be created through a continuous learning pattern of AI-based technologies \cite{latif2022ai}.}
    
    \item \textcolor{black}{\textbf{Effectiveness:} Blockchain-based technologies have a limited datastore capacity. When the data in this restricted database, which is essentially level DB, grows, the cost of reading and writing data increases as well. By utilizing AI-based approaches/algorithms, BC's storage methods may be further enhanced. As a result, it improves the efficiency of the Blockchain-based data management system by speeding up data queries \cite{van2022ai}.}
\end{itemize}

\subsection{Integration of AI-IIoT}
\subsubsection{\textcolor{black}{Discussion of Security Challenges}}
\begin{figure}[h]
    \centering
    \includegraphics[scale=0.29]{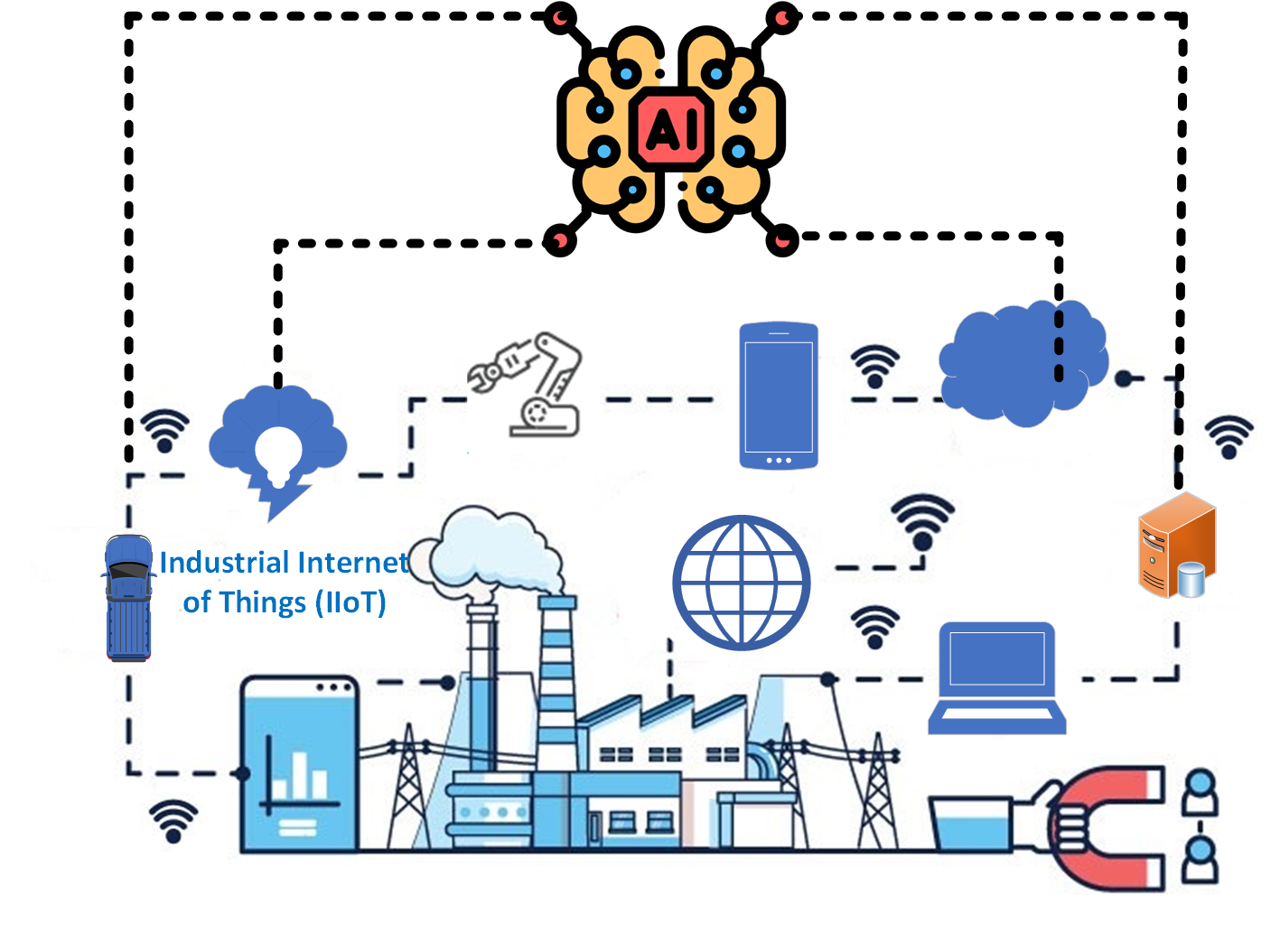}
   \caption{\textcolor{black}{AI with smart Industrial IoT Process}}
    \label{fig:f4}
\end{figure}

For an AI-enabled IIoT system that is shown in Fig. \ref{fig:f4}, additional security considerations for the AI system, its data, and its training ecosystem must be considered. Furthermore, the particular requirements imposed by IT/OT convergence, as well as the inter-connectivity of OT and IT concerns, must be addressed. Artificial intelligence (AI) is a data-driven technology. 
Some security concerns are listed below:
\begin{itemize}
    \item \textbf{Confidentiality: }AI works in a data-heavy environment. In general, confidentiality poses two issues that must be handled. For starters, there may be circumstances in which secret data cannot be mixed with non-confidential or other sensitive data, whether operational or training data. Second, it is critical to ensure that the insights produced by the AI system based on secret data are not utilized to infer the source data. Mechanisms for ensuring the confidentiality of business and commercial data must be put in place.
    
    \item \textbf{Explainability: }Understanding an AI system may be approached from two angles, both of which are related to the context of the AI system's employment inside the specific IIoT application. To begin, an IIoT system that will contain AI must comprehend how the AI system works, including its predictive nature, so that the total system can be specified, architected, and constructed. Second, in order to create trust in the system and its insights, the end-user of the AI-based IIoT service must comprehend what the AI component is doing.\newline
    
    \item \textbf{Controllability: }With AI-enabled IIoT systems, controllability is a critical security and safety risk. However, it has yet to be demonstrated that AI controllability will always be entirely achievable. This is due to the fact that the problem contains multiple subtypes, which might lead to inherent ambiguity.\newline
    \textbf{Explicit control: }AI can immediately interrupt the line, even if it is in the middle of an operation.\newline 
    \textbf{Implicit control: }AI attempts to stop the line at the first safe chance.\newline \textbf{Aligned control: }AI relies on its model of human intentions behind the order and employs common sense interpretation of the command to do what humans hope will happen.\newline \textbf{Delegated control: }If AI considers that the process could benefit from an improvement, it will stop it without waiting for a human to issue a directive.
    \item \textbf{Ethical and Societal Concerns: }The context of the application of AI technology raises a number of issues concerning ethics, societal concerns, bias, safety, the influence on labor, and policy in general. To avoid unforeseen outcomes, AI must be employed properly. A rising number of countries have established AI ethics standards and principles.\newline
    \textbf{Human, social, and environmental wellbeing: }AI systems should help individuals, society, and the environment throughout their existence.\newline
    \textbf{Human-centered values: }AI systems should respect human rights, diversity, and individual autonomy throughout their existence.\newline
    \textbf{Fairness: }AI systems should be inclusive and accessible throughout their lifecycle and should not include or result in unfair discrimination against individuals, communities, or groups.\newline
    \textbf{Privacy protection and security: }AI systems should respect and uphold privacy rights and data protection throughout their lifecycle, as well as ensuring data security.\newline
    \textbf{Reliability and safety: }AI systems should perform consistently in accordance with their intended purpose throughout their existence.\newline
    \textbf{Transparency and explainability: }Transparency and responsible disclosure should be implemented to guarantee that people are aware when an AI system is having a substantial impact on them and can determine when an AI system is interacting with them.\newline
    \textbf{ Contestability: }When an AI system has a substantial impact on a person, community, group, or environment, there should be a timely method in place to allow people to question the AI system's use or output.\newline
    \textbf{Accountability: }Those responsible for the various phases of the AI system life cycle should be identified and held accountable for the AI systems' outputs, and human supervision of AI systems should be enabled.\newline
\end{itemize}

It is critical to understand the following concepts while creating an AI system and incorporating it into an IIoT system:
\begin{itemize}
    \item which assets must be protected,
    \item what are the corresponding data governance models,
    \item what measures are required to protect the security of AI systems and
    \item what are the broader data protection considerations in regard to AI, such as institutional and policy regulations for data privacy and confidentiality.
\end{itemize}
The general security design principles that guide the capabilities and tactics used in implementing security in IIoT systems (based on Saltzer and Schroeder \cite{raj2021establishing}) must also be applied to AI security throughout implementation:
\begin{itemize}
    \item \textbf{Principle of the economy of mechanism: }The principle of mechanism economy is to keep the design as simple and minimal as feasible.
     \item \textbf{Principle of fail-safe defaults: }The principle of fail-safe defaults states that access decisions should be based on permission rather than exclusion.
      \item \textbf{Principle of complete mediation: }The principle of comprehensive mediation states that access to all objects must be checked for authority.
       \item \textbf{Principle of open design: }procedures should not rely on possible attackers' ignorance, but rather on the possession of explicit and easily safeguarded keys.
        \item \textbf{Principle of separation of privilege: }Wherever possible, a protective mechanism that requires two keys to open it is more resilient and adaptable than one that simply provides access to the presenter of a single key.
         \item \textbf{Principle of least privilege: }The principle of least privilege states that every application and every user of the system should utilize the fewest set of privileges necessary to fulfill the task.
          \item \textbf{Principle of least common mechanism: }The principle of least common mechanism states that the amount of mechanism that is shared by more than one user and relied on by all users should be kept to a minimum.
          \item \textbf{Principle of psychological acceptability: }The psychological acceptability principle states that the human interface should be developed for ease of use so that users appropriately implement the protection measures on a regular and automatic basis.
\end{itemize}

\subsubsection{Integration Benefits}

\begin{itemize}
    \item \textcolor{black}{\textbf{Data and Device Management:} The IIoT is primarily concerned with the management and efficacy of the OT. OT is a broad term that encompasses both hardware and software that regulates the physical system's functioning. The use of AI and IIoT in the management of OT can be beneficial.}
    
    \item \textcolor{black}{\textbf{Data Security:} The information created by IIoT activities is sensitive and private. If these data are not securely protected, they might possibly lead to the system's demise. Furthermore, analyzing such a large volume of data is tough. In addition, intrusion detection using an Artificial Neural Network (ANN) and data analysis beneficiary for the IIoT industry \cite{kanimozhi2019artificial}. }
    
    \item \textcolor{black}{\textbf{Data Analysis:} Data analytic is the process of analyzing data based on the data's history and the system's continual assessment of the data it receives. The AI-based intelligent system allows for the prediction of hazards as well as potential advantages\cite{de2022ai}. }
    
    \item \textcolor{black}{\textbf{Decision-Making Strategy:} The most recent Industrial IoT trend is communication-based, with some technology integration and data. However, data collection alone will not solve an IIoT design problem. It has a decision-making component that supports the development of an intelligent system for connected devices. This issue incorporates environmental control, horticulture, water management, energy management, and other components \cite{EasyChair:5023}. In today's IR 4.0 environment, improved AI-based technologies have made the operations involved in the factory production chain, from manufacturing to raw sample collection, more efficient. Tasks that need human involvement may be assigned to robots with the use of AI blessings \cite{ahmed2022artificial, rahman2023towards}.}
    
\end{itemize}

\subsection{Integration of BC-IIoT}

\subsubsection{Discussion of Security Issues}
\begin{figure}[h]
    \centering
    \includegraphics[scale=0.43]{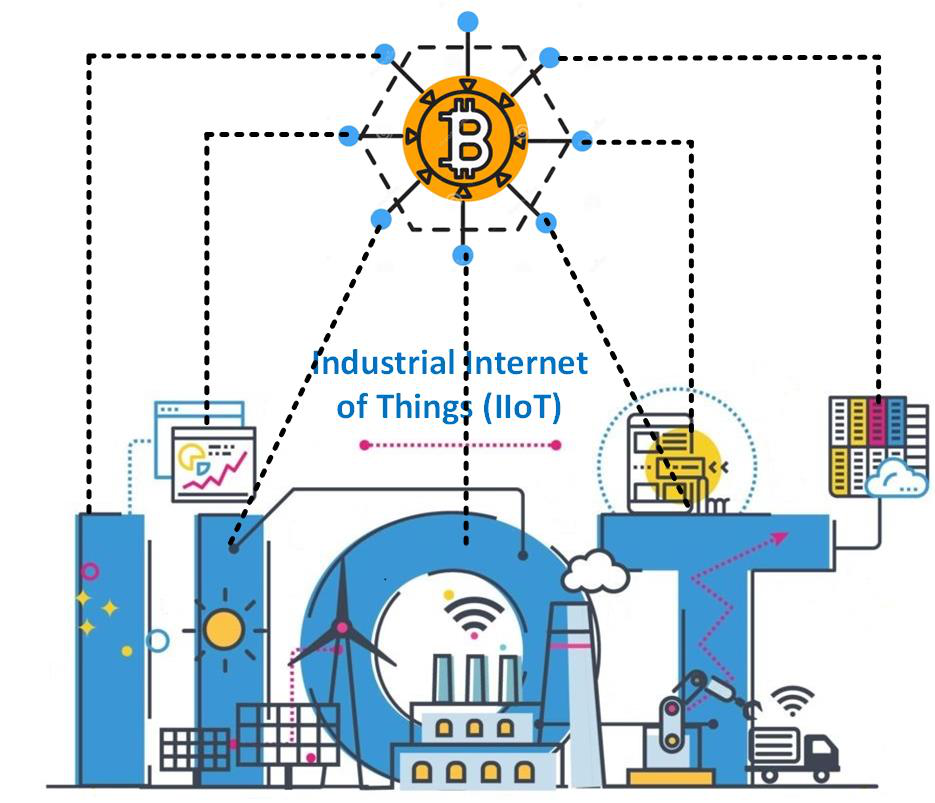}
   \caption{\textcolor{black}{Blockchain with smart IIoT Scenario}}
    \label{fig:f5}
\end{figure}

\textcolor{black}{The combination of IoT and Blockchain opens up a slew of new possibilities for Industry 4.0 \cite{wang2020blockchain}. The combination of them is depicted in Fig. \ref{fig:f5}. However, there are a number of issues that must be addressed before the Blockchain-assisted IoT or IIoT methods can be released. Discussions have been made on a few of the most important open research challenges so far.}

\begin{itemize}

    \item \textcolor{black}{Devices with Resource Constraints: The devices utilized in IoT and IIoT systems are typically resource-constrained. These devices have limitations in terms of processing power and storage capacity. As a result, they are not suited for deployment on the Blockchain network since they require a lot of computing power and storage space \cite{reyna2018blockchain}.
    }
    
\textcolor{black}{
    \item An Appropriate Incentive Mechanism for a Blockchain-Powered IoT System: The basic functionality of Blockchain technology is the incentive system. Because solving the problems demands a lot of computational power, the mechanism depending upon proof-of-work is unsuitable for the IoT system. After mining 210, 000 blocks, the block reward is also halved \cite{saito2019make}.
    As a result, determining an incentive mechanism that is appropriate for an IoT or IIoT system is an area of investigation for future research.}
    \textcolor{black}{
    \item Big Data Analytics Scheme: The Internet of Things is commonly utilized for data-related applications in which the devices generate a large volume of multimedia data\cite{vyas2019converging}. The obtained data is saved on a cloud server, where the machine learning algorithm performs data acquisition, extraction, and analysis.}
    \textcolor{black}{
    \item Scalability of a Blockchain-Enabled IoT System: The existing Blockchain mechanism's scalability also restricts the usability of IoT and IIoT systems. IoT device-based applications (such as smart healthcare) generate a large amount of data quickly. As a result, current platforms aren't directly applicable to IoT or IIoT systems.}
    \textcolor{black}{
    \item Security Threat and Privacy Breach: The implementation of the Blockchain method in the area of IoT undoubtedly increases the defense of IoT systems by the use of the fundamental mechanism of Blockchain, which is encryption and hashing. However, the components are linked through a wireless communication system (mostly). However, this approach is not secure because it can be cracked using a machine learning algorithm\cite{conti2018survey}.}
    \textcolor{black}{
    \item IoT requires Federated Learning (FL): The conventional learning technique cannot be employed in the IIoT due to scattered and private datasets. FL is a system in which devices execute locally trained data and then transfer the results to the server for model aggregation on a global scale. This is a novel research direction, and in order to combine Blockchain-assisted FL for resource-limited IoT equipment's, extensive analysis is necessary\cite{rahman2023federated, khan2021federated}.
    }
\end{itemize}

\subsubsection{Integration Benefits} 
\textcolor{black}{Industrial IoT systems, as stated in the previous subsection, have numerous challenges, including heterogeneity, poor interoperability, and device resource constraints. To address these issues, Blockchain technology can be used in conjunction with current IIoT platforms. When compared to standard IIoT platforms, integrating Blockchain into IIoT platforms offers various potential benefits.}
\begin{itemize}

    \item \textcolor{black}{Enhanced Interoperability: Blockchain might potentially increase the interoperability of IIoT platforms by converting and storing data records in a shared Blockchain. To achieve interoperability, Blockchain-enabled IIoT platforms must maintain some public information from their equivalents, such as authorization rules, user-associated public keys, and audit logs of data access. This has the potential to greatly increase interoperability in IIoT applications.
    }
    \textcolor{black}{
    \item Improved Security: Blockchain can deliver safety improving solutions due to the Blockchain's inherent security qualities, such as confidentiality and availability. Because all legitimate data are saved as Blockchain settlements which are encoded and digitally signed, the IIoT data will be secure. Under Blockchain-enabled authentication, this procedure assures that all interactions with the IIoT platform stay private.}
    \textcolor{black}{
    \item Greater Transparency: Blockchain technology improves data and transactional data exchange transparency. As a distributed and decentralized ledger, all network members have the similar data in their own versions, which may only be altered through consensus methods. Any changes to a single transaction would necessitate changes to all future records, which could necessitate network-wide collusion\cite{francisco2018supply}.
    }
    \textcolor{black}{
    \item Improved Traceability: The Blockchain opens up the possibility of resolving major traceability issues that plague traditional IIoT platforms. The importance of traceability in the verification of industrial transactions across different industries cannot be overstated.
    }
    \textcolor{black}{
    \item Improved Corporation: Using Blockchain, corporate cloud service providers can connect with IIoT consumers without the need for a central authority. Even in an untrustworthy setting, IIoT data is safely sent using Blockchain management.
    }
\end{itemize}

\subsubsection{BC, and IR 4.0 for Security Purposes}
In the advanced world with the development of IoT, the industrial sectors now established industry 4.0 applications that use a wide number of sensors and devices. These devices communicate with each other and transmit data all around the globe. To deal with uninterrupted and secured data transmission, some issues must be ensured, such as data transparency, heterogeneity, privacy, redundancy, security, and so on \cite{rahman2021smartblock}. Moreover, the applications within the IR 4.0 ecosystem need to get access to data servers that may be on different locations with a not similar format. There are some known attacks that may hamper the industrial work lie, DDoS, network routing congestion, manipulation of data or phishing attacks are some attacks that can cause delay or harm to the data transferred within the network. With the increase of automation in the era of IR 4.0 the ratio of vulnerabilities also increased \cite{9528133}. \vspace{2mm}

Blockchain is one of the powerful technologies with cryptographic capabilities that help to ensure the security of IR 4.0 applications. It prevents unauthorized access by imposing security measures like a private key Blockchain that is used to sign the digitally passed message, and on the other hand, a public key helps to verify. Further, the hashing mechanism helps to implement data integrity and also ensures that all of the transactions are stored in a safe block, and those blocks cannot be manipulated without the matching hash value \cite{dasgupta2019survey}. Again, for the base station, Blockchain allows authentic boot, counter, firmware, and encryption for data \cite{minoli2018blockchain}. After that, a stable way of communication is allowed in the top layers for the public key system and cloud architecture Blockchain \cite{nagasubramanian2020securing}. Decentralization, the authentication of centralized structure to decentralized management; thus, the authentication in a cost-efficient way is possible by incorporating Blockchain mechanism \cite{noh2019study}. Lastly, because of the cryptographic feature of the technology, security and privacy are preserved in Blockchain technology. Furthermore, The software-defined network is another solution to the security challenges of IR 4.0 \cite{Rahman2020}. Further, the SDN controller is considered as the brain of the model interface with detection and prevention of intrusion through the northbound API \cite{9350419, el2020sdn}.

\vspace{2mm}
\subsubsection{Industry 4.0 Applications Challenges and Services}
\begin{figure}[h]
    \centering
    \includegraphics[scale=0.28]{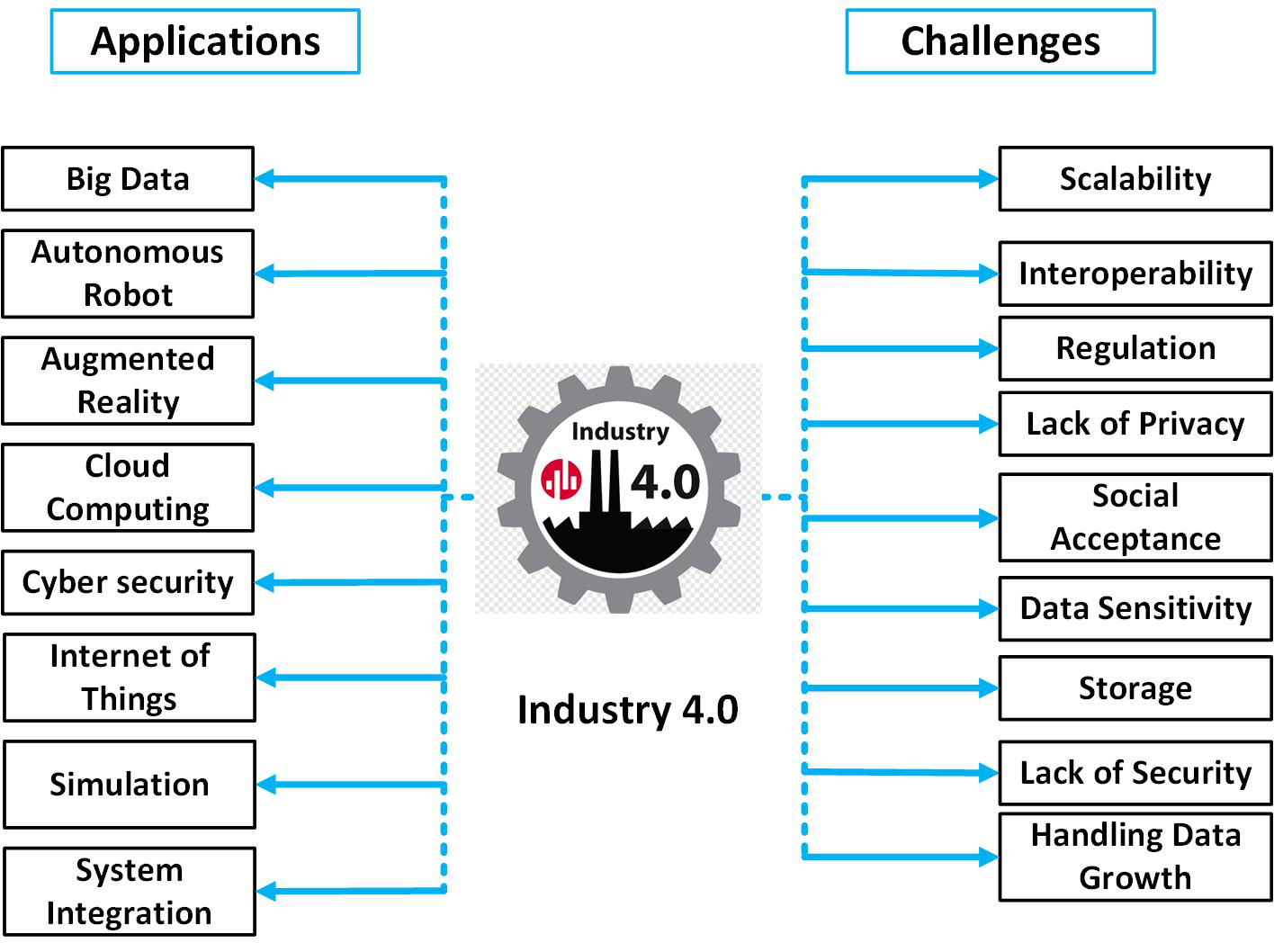}   \caption{\textcolor{black}{Applications and Challenges Industry 4.0}}
    \label{fig:f6}
\end{figure}

The application of IR 4.0 is countless, from medical to education, manufacturing all sectors are now implementing the features of IR 4.0 to achieve the goal of automation. During the present pandemic situation of COVID-19, the application of IR 4.0 is huge, which is shown in Fig. \ref{fig:f6}. To detect or predict any disease, scan temperature, or other symptoms that can be easily observed by the outside bodily criteria, distribution of vaccine \cite{9641303}, or primitive measures for healthcare workers or professionals, or providing services during lockdown possible through the application of industry 4.0 applications \cite{javaid2020industry}. In the industrial sector for the manufacturing sector, developing intelligent products, collecting important data for the improvement of the design process, assessment of products, and management of prototypes are possible with the IR 4.0 applications \cite{miranda2019sensing, bressanelli2018role, goodall2019data}. The security issues of IR 4.0 can further be subdivided into the layers of the architecture. The layer that holds the smart devices for sensing purposes faces attacks related problems like trying to get access from unauthorized users, maintaining the confidentiality of the user, and handling noisy data. Then the layer that is responsible for networking purposes needs to handle routing jams or congestion during data transmission. The application layer that manages the application faces security attacks such as, phishing attacks or malicious code attacks. Then, the service layer responsible for the management of data servers is the main attraction of the intruders by attacking the layer for example DoS attack, data manipulation, and so on \cite{kumar2020challenges}. 

\subsubsection{Solutions to Address the IR4.0 Challenges}
The base of Industry 4.0 is IoT, and it faces many challenges like attacks in the IoT ecosystem by a third party or data transparency, or providing security and privacy to the applications under Industry 4.0. To overcome the challenges of IR 4.0, some additional technology has been attracted to researchers. For example, Blockchain technology, which is well known for its consensus mechanism and encryption procedure, helps to protect data from being manipulated by intruders by creating a secured chain-like structure. Further, it provides a transparent method for the user\cite{gupta2020blockchain}. Again, the machine learning that is used for better prediction can be used to train a system and then predict the possibility of attacks under certain circumstances. The SDN is able to divide the overall network by leveraging its data, control, and application plane and also operates with the switches that send data to and forth. The controllers control the overall process. It is clear that SDN creates a centralized program. Through this centralized medium, the mitigation of attacks is easier, and thus, the system protects the networking system \cite{wang2021mitigating}. Moreover, the establishment of wireless communication is another challenge in the edge of IR 4.0 to achieve automation flawlessly. In this regard, 5G technology is a possible solution that can connect an array of communicating devices with the option of flexibility and efficiency. Capable of handling a large number of IoT devices \cite{rao2018impact}. To sum up this section, we have presented a Table \ref{tab:T4} that has discussed the previous studies based on the integration and motivation of the BC, AI, and smart Industrial IoT.

\subsection{\textcolor{black}{Key Comparative Metrics Analysis based on BC, AI, and smart IIoT}}
\textcolor{black}{In table \ref{tab:T2}, the performance metrics that have been used to analyze the performance under different scenarios is depicted. Here, we have used different metrices related to AI, BC and smart IIoT field. In addition, we have mentioned the application areas of the recent individual studies which have been used in Artificial Intelligence, Blockchain, and smart Industrial IoT technology. Moreover, we have analyzed the comparative study, including the analysis of these metrics related to different technologies, which has included individual application areas}.

\begin{table*}
    \centering
    \scriptsize
        \caption{\textcolor{black}{Performance Metrics analysis based on BC, AI, and IIoT technologies}}
    \begin{tabular}{|p{0.7cm}|p{0.2cm}|p{0.3cm}|p{0.4cm}|p{1.2cm}|p{0.8cm}|p{1.4cm}|p{1.7cm}|p{0.6cm}|p{0.7cm}|p{1.0cm}|p{1.2cm}|p{1.5cm}|}
    \hline
        \multicolumn{1}{|c}{\multirow{2}{*}{Study}} & \multicolumn{3}{|c|}{Technologies Used} & \multicolumn{8}{|c|}{Performance Metrics} &
       \multicolumn{1}{c|}{\multirow{2}{*}{Applications Areas}} \\
         \cline{2-12}
          & BC & AI & IIoT & Throughput & Response Time & Communication Overhead & Bandwidth and Latency & Energy Consumption & Security \& Privacy & GAS Consumption & Block Validation Time &  \\ 
         \hline

        Bu et al. \cite{bu2021iiot}& X &\checkmark& \checkmark& X & \checkmark & X & X& \checkmark & X & X & X & \textcolor{black}{Manufacturing Process}\\
        \hline
        Sasikumar et al. \cite{sasikumar2022sustainable} &  \checkmark & \checkmark & \checkmark & X & \checkmark & X & X & \checkmark & \checkmark & X & X &\textcolor{black} {Secured system for IIoT} \\
        \hline
        Salim et al. \cite{salim2022blockchain} & \checkmark & X & \checkmark & X &  \checkmark& X &  \checkmark & \checkmark & \checkmark & X & \checkmark & \textcolor{black} {BC for IIoT secuirty by eraly detection of BotNet}\\
        \hline
        Yazdinejad et al. \cite{yazdinejad2022block} & \checkmark & \checkmark & \checkmark & \checkmark & X & X & \checkmark & X & \checkmark & X & X & \textcolor{black}{Cyber threat finding thorugh FL and BC for Smart IIoT}\\
        \hline
        Li et al. \cite{li2022blockchain}& \checkmark & X & \checkmark & \checkmark& X & X & \checkmark & X& \checkmark & X & X & \textcolor{black}{ BC based supply chain management.}\\
        \hline
        Alrubei et al. \cite{alrubei2022secure} & \checkmark & \checkmark & X & \checkmark & \checkmark & \checkmark & \checkmark & \checkmark & X & \checkmark&X &\textcolor{black}{Support to the IoT system.}\\
        \hline

        Ratta et al. \cite{ratta2021application} & \checkmark & \checkmark & X & \checkmark & X & X & X & X & \checkmark & X&X &\textcolor{black}{BC in medical sector security.}\\
        \hline
        Koushik et al. \cite{koushik2019performance} & \checkmark & \checkmark & X & \checkmark & X & X & X & X & \checkmark & X&X &\textcolor{black}{BC in medical sector security.}\\
        \hline
        
        Ismail et al. \cite{ismail2020performance} & \checkmark & X & X & \checkmark & X & X & X & X & \checkmark & X&X &\textcolor{black}{BC in medical sector security.}\\
        \hline
        Roehrs et al. \cite{roehrs2019analyzing} & \checkmark & X & X & \checkmark & X & X & X & X & \checkmark & X&X &\textcolor{black}{BC in medical sector security.}\\
        \hline

\end{tabular}
    \label{tab:T2}
\end{table*}

\begin{table*}[h!]
\scriptsize
\caption{Existing Studies Analysis based on Applications of AI-BC with IIoT}
 \label{tab:T5}
\centering
\begin{tabular}{ |p{2cm}|p{4cm} |p{4cm}|p{6cm}|}
\hline
\textbf{Authors \& Year} & \textbf{Applications Areas} & \textbf{Used Techniques \& Tools} & \textbf{Major Drawbacks} \vspace{1mm} \\ 
\hline
Khezr et al. \cite{khezr2022edge}(2022) & Data management in IIoT era & Blockchain, IIoT, Edge Computing  & \textcolor{black}{Comparison with other consensus approach not present.}\\
\hline
Kumar et al. \cite{kumar2022artificial} (2022) & Business & Blockchain, AI & \textcolor{black}{Attack detection and mitigation process requires further analysis.}\\
\hline
Lin et al. \cite{lin2022intelligent} (2022) & Large scale industry  & Blockchain, IIoT , AI & \textcolor{black}{Complexity cost of implementing such method not analysed vastly.} \\
\hline
Rizwan et al. \cite {rizwan2022simulation} (2022) & Machine learning, Blockchain, IIoT, VANET & Traffic management & \textcolor{black}{Possibilities of attacks in the case of data transmission not considered in this work.}\\
\hline
Bhargava et al. \cite{bhargava2022industrial} (2022) & Management of supply chain & AI, IIoT, VANET & \textcolor{black}{Complexity and cost of implementing such method in real world scenario requires further study.}\\
\hline
Rajawat et al.\cite{rajawat2022renewable} (2022) & IIoT, AI & IIoT energy management & \textcolor{black}{Privacy and security ensuring with such methods are challenging.}\\
\hline
Wang et al. \cite{wang2022blockchain} (2022) & Recourse trade system & Blockchain, Edge Computing, IIoT & \textcolor{black}{Workload forecasting  not taken into account.} \\
\hline
\textcolor{black}{Safa et al. \cite{safa2022enhancing} (2022)} & Supply chain management & Blockchain, IIoT & \textcolor{black}{Analysis of cost to implementation such method not present in this work.} \\
\hline

Wang et al. \cite{wang2021enabling} (2021) & IIoT applications & Blockchain, IIoT, Transfer Learning & \textcolor{black}{Comparison with other related work not presented.}\\
\hline

Kalinin et al. \cite{kalinin2021ai} (2021) & Transportation system &  Blockchain, AI, IIoT, VANET & \textcolor{black}{Cost and effectiveness of such method with real time data requires analysis.}\\
\hline

Yu et al. \cite{yu2021blockchain} (2021) & To authenticate user data in the IIoT based application &	Blockchain, IIoT & \textcolor{black}{The user in this system are fully rescinded and another issue in this work is the number of transactions shown per second is between 100 to 1000.}.\\
\hline
Zhu et al. \cite{zhu2021green} (2021) & Industry & AI, Edge Computing, IIoT & \textcolor{black}{The dataset used is not real time dataset.}\\
\hline

Luo et al. \cite{luo2021great} (2021) & Business perspectives & Deep learning, Blockchain & \textcolor{black}{The simulation based result does not include real world scenario.}\\
\hline

Kumar et al. \cite{kumar2021iiot} (2021) & Indian small along with medium sized organizations & Blockchain, IIoT & \textcolor{black}{The paper focuses on the small and medium enterprises excluding large sectors.}\\
\hline
Rahman et al. \cite{rahman2021ai} (2021) & Smart City & AI, IIoT, CPS, Deep learning & \textcolor{black}{In real world based implementation scenario the results not tested.}\\
\hline

Jogunola et al. \cite{jogunola2020consensus} (2020) & Energy based trading organization & Blockchain, Reinforcement learning, IIoT, & \textcolor{black}{The integration benefits of such technologies not presented.}\\
\hline

Rahman et al. \cite{rahman2020distb} (2020) & Industry 4.0 application & Blockchain, IIoT & \textcolor{black}{Comparison with other related research work is not analyzed to a great extend.}\\
\hline

Wu et al. \cite{wu2020convergence} (2020) & Industry 4.0 & Blockchain, IIoT & \textcolor{black}{Complexity analysis of implementing such methods in real world scenario requires further analysis.}\\
\hline

Wang et al. \cite{wang2020reinforcement} (2020)  & Transportation management & Reinforcement learning, IIoT & \textcolor{black}{In worst case scenario like, delay time the system not tested.}\\
\hline

Wazid et al. \cite{wazid2020private} (2020) & Healthcare system & Blockchain, AI, Drone technology & \textcolor{black}{Using BC for the peer to peer connectivity requires massive data; thus there might be some security challenges.}\\
\hline
\end{tabular}
\end{table*}

\section{\textcolor{black}{Individual \& Collaborative Applications: BC, AI, IIoT}}\label{sec:tacol}

\subsection{Individual Applications Overviews}

\begin{figure}[h]
    \centering
    \includegraphics[scale=0.28]{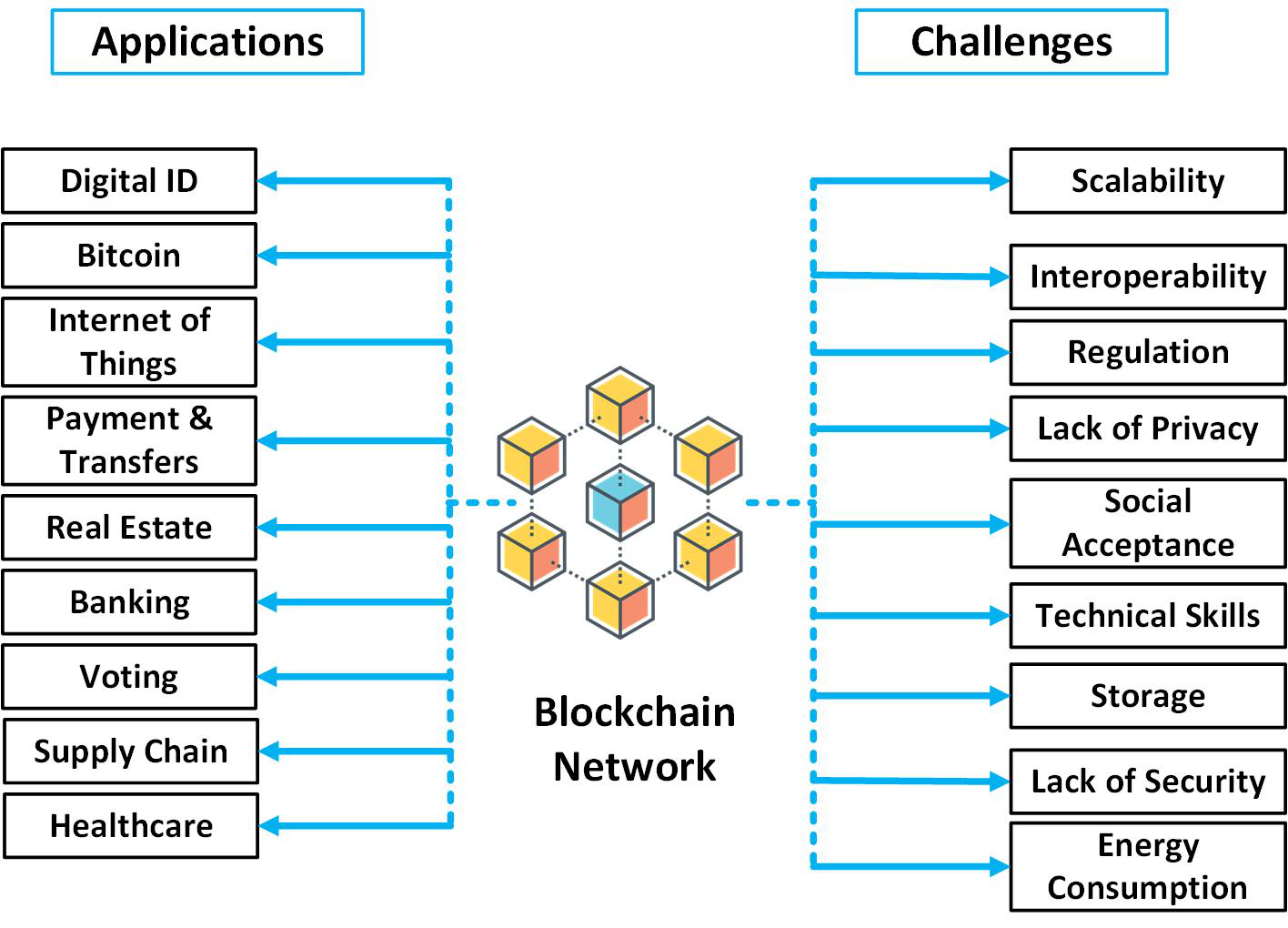}
   \caption{\textcolor{black}{Scenario of Blockchain Applications and Challenges}}
    \label{fig:f7}
\end{figure}

\textcolor{black}{Blockchain technology plays a critical role by ensuring data security while being exchanged. Further, in a smart building or city management system, data transfer is a must in order to construct a smart system \cite{rahman2020distblockbuilding}. It is impossible to make any decisions or move forward without accurate knowledge about a single component of a system. The creation of any energy, system, or object is considered by gathering and processing user needs. As a result, data exchange is more common, and data security is a primary consideration in these situations. Because, every single lack of knowledge or change of data will result in fictional output outcomes that are of no service to the world. The cryptographic method offers a better approach to safeguard data from hackers by encrypting it and storing it in a chain of blocks that is difficult to modify by a third party. In this way, other smart systems like smart agriculture, smart grid, and smart vehicular system are all the base of the data communication system. As a result of securing these data or information, the application of Blockchain in remarkable \cite{abdelmaboud2022blockchain}. The application of BC technology has been visualized in Fig. \ref{fig:f7}. Again,  data collecting alone cannot be the solution to a smart city's design dilemma. It has a decision-making portion that aids in the creation of an intelligent system for linked devices. The environment control, horticulture, water management, energy management, and other components are incorporated in this situation. AI is also being used in the field of parking and traffic management, which is becoming more important as the number of vehicles on the road increases, giving a better answer for a country's rising population. Then, using AI, a drone-based surveillance system, door-to-door service for senior people, and effective monitoring are all achievable in the smart cities scenario. The procedures involved in the factory production system, from manufacturing to raw sample collecting, have become efficient in today's current IR 4.0 situation by employing advanced AI-based approaches. Smart devices, sensors, and sophisticated AI algorithms are all used to create intelligent factory management system. Again, with the use of AI blessings, tasks that require human involvement may be delegated to robots \cite{ahmed2022artificial,Islam_Rahman_Kabir_Khatun_Pritom_Chowdhury_2021}.}

\begin{figure}[h]
    \centering
    \includegraphics[scale=0.27]{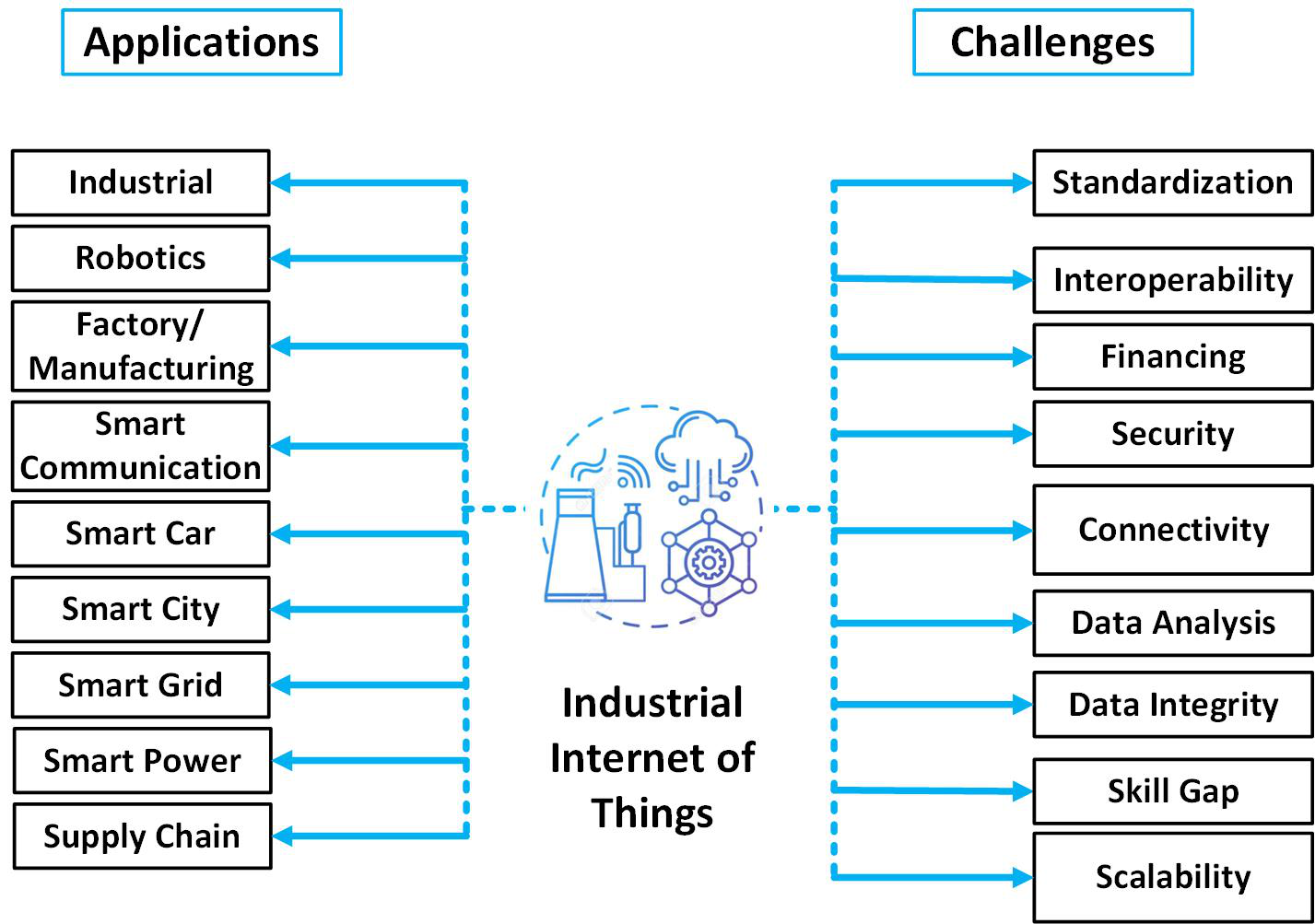}
   \caption{Smart Industrial Internet of Things Applications and Challenges}
    \label{fig:f67}
\end{figure}

\subsection{\textcolor{black}{Collaborative Applications Overviews}}
\subsubsection{BC-AI Applications}
\begin{itemize}
    \item \textcolor{black}{\textbf{Data management with BC-AI:} By blending a decentralized Blockchain with an intelligent way that is AI-enabled, many smart systems can securely communicate with a large number of agents from various suppliers. For example, Wang et al. \cite{wang2019securing}  presented a solution for trustworthy data exchange that integrates Blockchain and AI by introducing intelligent security rules. In another similar research, Bertino et al. \cite{bertino2019data} emphasized the necessity of guaranteeing data openness in an algorithmic manner, as well as a Blockchain-based technique for data security in AI-based systems.}
    
    
    \item \textcolor{black}{\textbf{Decentralized Intelligence:} The rapid proliferation of IoT has resulted in a massive amount of IoT data. We can get learning outcomes and models from the vast IoT data using the AI service. Due to the wide deployment of IoT and edge computing devices, coordination with several devices is usually required to complete difficult model training tasks. There are two ways to collaborate in this situation. For complete data analysis and prediction, distinct IoT devices or edge devices must first communicate data (which includes smart monitoring, monitoring in several areas require to share information).\newline
    LearningChain, a decentralized security learning mechanism, was suggested in \cite{chen2018machine}. There are two types of members in the learning chain network: data holders and compute nodes. The data holder pays for the computational node's assistance in training the learning model. Finally, several data sources collaborate to train a global model.}
    
    \item \textcolor{black}{\textbf{EHR Data management with BC-AI:} In a smart healthcare system, the integration of different smart technologies and the transfer of data have increased vastly, which implies the importance of securing data in the healthcare sector as it comprises of sensitive user information \cite{bhuiyan2023iot}. Chamola et al. \cite{chamola2022artificial}, emphasized the significance of AI in the event of a medical emergency since, via this method, hospital administration and doctors may rapidly retrieve the stored information about a patient. In this study, AI (OCR technique) assists the staff in giving stored information in a very short period of time while maintaining Blockchain-based data security. With the advancement of intelligent agent-based robotics, the use of robots in surgery has grown in popularity. However, the use of robots in this role raises concerns about security. In this context, Gupta et al. \cite{gupta2020bats} presented an AI technique based on the XGBoost algorithm with Blockchain integration for data security and privacy. According to Tagde et al.  \cite{tagde2021blockchain}, the Blockchain-based technique supplements the AI-based EHR data storage mechanism, in which AI maintains system security and Blockchain cryptographically handles EHR data.}
    \item \textcolor{black}{\textbf{Internet Security Enhancement:} In today's modern world, internet usage has skyrocketed, resulting in an increase in human-machine or machine-machine contact. Approximately 52\% of traffic on the web subsystem is by the bots; the necessity of previous history records is a must for future communication between them to reduce traffic. Progga et al. \cite{progga2021securing} proposed to solve the VANET's security issues with Blockchain and AI with the smart network management system.}   
\end{itemize}

\subsubsection{AI-BC-IIoT Applications}
\textcolor{black}{ At the moment, the application of IIoT has grown significantly, including data storage management and data transmission, in order to achieve the objective of making an industry smart. The integration of the cloud has accelerated the growth of IIoT, but it has also caused several issues in terms of privacy and security of information storage, deletion, or upgrade. In order to make IIoT applications more secure and private, researchers proposed solutions. Among numerous technologies, BC is one of the most promising for ensuring the security of user data through a cryptographic solution. }
\textcolor{black}{
\begin{itemize}
    \item \textcolor{black}{\textbf{Smart Manufacturing:} The smart manufacturing industry is a broad category of manufacturing that includes cloud, IoT-enabled technology, and service-oriented manufacturing, among other technologies. These technologies, when combined, transform standard automated manufacturing into smart manufacturing. Most existing solutions, on the other hand, pursue a consolidated industrial structure and rely on a intermediary administration. Blockchain-AI-powered IIoT platforms typically address interoperability difficulties by joining several IIoT systems via a P2P network and permitting information exchange over various industrial areas.}\newline
    \item \textcolor{black}{\textbf{Smart Grid:} With rising energy demands to provide facilitate production activities, smart energy and associated systems are becoming increasingly important in several industrial environment. The rise of distributed green energy sources is transforming consumers' roles into prosumers, where an individual can participate in both generation and consumption of energy. The advent of BC-AI technologies opens up new possibilities for assuring secure peer-to-peer energy trade; additionally, some recent research has recommended using Blockchain to address difficulties in energy management systems (EMS).}\newline
    \item \textcolor{black}{\textbf{Supply Chain:} An industrial product is usually the result of collaboration between several suppliers from various production areas. Certain fabricated parts may, nevertheless, find their way into the supply chain. Systems which are able to apprehend fraudulent activities are fairly costly. This issue may be resolved by combining Blockchain with IIoT. When a part is formed, it is usually assigned a unique ID. This ID is then assigned an immutable timestamp. As tamper-resistant proof, each part's identity information can be recorded in a Blockchain. After-sale services in supply chain management can also be reduced with BC-AI-enabled applications.}\newline
    \item \textcolor{black}{\textbf{Testing and Validation:} The purpose of the testing and validation technique is to establish a set of KPIs for each conceivable use case and to assess the performance of our proposed AI-based strategy through trial and error. Because not only technical needs but also commercial feasibility and impact, as well as ethical considerations associated to the introduction of novel AI concepts and models, the evaluation plan incorporates technical, business, and ethical validation metrics.}\newline
\end{itemize}}
Furthermore, we analyze a table \ref{tab:T5} to describe different applications based on the considered technologies. The following are some of the case studies from the underlying BC-AI enabled Industrial IIoT research such as-- Yu et al. \cite{yu2021blockchain} proposed a solution based on the integration of Blockchain and IIoT, Where Blockchain is used to authenticate identity. The Blockchain is in charge of storing all keys and characteristics. In this suggested solution, the mechanism of intruder detection is used at each level to improve system security while using minimal overhead time. The factories may safely transport data within their network by utilizing this method. In similar study, Christidis et al. \cite{christidis2016blockchains} describe a smart contract and BC-based approach for automatic firmware updates. Decentralized BC-based intelligent production platforms can provide more secured and confidential safety than centralized structures. Again, Kim et al. \cite{kim2018toward} provided a track-ability reasoning based on the Ethereum Blockchain platform that integrates Blockchain, AI and IoT technologies to give tamper-proof evidence for products. Then, Xu et al. \cite{xu2017intelligent} presented a Blockchain-based crowd-sourced energy system (CES) that allows for P2P energy exchange at the distribution level, where resource holders can transact with each other devoid of assistance from trustworthy entities. Kumar et al. \cite{kumar2021iiot}, identified 10 flaws where the IIoT system may fail to offer security and proposed strategies to establish a system secret and safe for the Indian SMEs sector utilizing Blockchain-based techniques. Another syudy, Jogunola et al. \cite{jogunola2020consensus} analyzed the use of  Blockchain, Reinforcement learning, IIoT,  in the field of the energy-based trading organization. Moreover, in this article, the application of different consensus protocols that are mainly for Blockchain and AI has also been analyzed.

\section{\textcolor{black}{Open Issues and Opportunities}}
\label{sec:opissfo}

\subsection{Discussion of Open Challenges}
\textcolor{black}{The combination of IoT and Blockchain opens up a slew of new possibilities for Industry 4.0 \cite{wang2020blockchain}. However, there are a number of issues that must be addressed before the entire Blockchain-enabled IoT or IIoT system can be released. We've discussed a few of the most important open research problems so far.
\begin{itemize}
    \item \textbf{Devices with Resource Constraints:} Most of the devices utilized in IoT and IIoT systems are resource-constrained. As a result, they are not suited for deployment on the Blockchain network since they demand a lot of computing power and storage space. Furthermore, consensus algorithms such as proof of work take a significant quantity of energy and power to mine\cite{reyna2018blockchain, 9499121}. Therefore, the method dependent on this variety is unsuited for a resource-bound method.
    \item \textbf{An Appropriate Incentive Mechanism for an AI-BC-Enabled IoT System:} The basic component of AI-based Blockchain technology is the incentive system. Because solving the problems demands a lot of processing power, the incentive process dependent on proof-of-work is unsuitable for the IoT system. After mining blocks, the block reward is also halved\cite{saito2019make}. As a result, determining an incentive mechanism that is appropriate for an IoT or IIoT system is an exploration area for scholars.
    \item \textbf{Scheme for Big Data Analytics:} The Internet of Things (IoT) is commonly utilized for data-related applications in which devices create large amounts of multimedia data\cite{vyas2019converging}. The created data is saved on a cloud server, where the machine learning algorithm performs data preparation, extraction, and analysis. Important concerns with this method include privacy leaks, access control measures, and so on.
    \item \textbf{AI-Blockchain-Enabled IoT System Scalability:} The current Blockchain technology's scalability further bounds the suitability of IoT and IIoT systems. The bitcoin technology has a lower transaction throughput (5 to 7 transactions per second, with a 10-minute average block time), whereas Ethereum has a 15-second average block time\cite{croman2016scaling}. On the other hand, IoT device-based applications (such as smart healthcare) quickly generate a large amount of data. As a result, existing platforms aren't immediately applicable to IoT or IIoT systems.
    \item \textbf{Decentralized IoT frameworks using machine learning and deep learning:} With the improvements in AI and machine learning over the last several years, we've seen dramatic changes. In IIoT networks, machine learning (ML) and, in particular, deep learning (DL) algorithms may be used to make intelligent decisions. In IIoT networks, it may optimize a variety of industrial activities and energy trading procedures. The use of AI in conjunction with the IIoT can help to maintain trust across various sources and network members. IIoT users will be able to monetize their data and crowdsource data to ML models for IoT services once a BC-enabled system is implemented. In addition, combining ML and DL methods in a Blockchain-based system can improve the security and performance of IIoT systems in several ways.
    \item \textbf{Framework for Dynamic Security:} Heterogeneous devices, ranging from low-power devices to high-power servers, are connected in the IoT system. As a result, a single solution may not be applicable to all Blockchain-enabled IoT systems. Furthermore, the security solution should take into account the nature of resource-constrained IoT devices, as well as the end customers' basic security requirements\cite{novo2018blockchain}. As a result, developing flexible and adaptive defense architecture for a Blockchain-assisted IoT (or IIoT) technology is an exciting investigation topic.
    \item \textbf{Vulnerability in Security and Privacy Leak:} Through the fundamental mechanism of Blockchain, which is encryption and hashing, the use of Blockchain technology in the realm of IoT unquestionably increases the safety of IoT systems. However, the devices are (primarily) connected by wireless communication systems, which are vulnerable to protection gaps, which include replying attacks, overhearing, and jamming owing to the open wireless channel.
    \item \textbf{Transaction Rejection:} Edge nodes, especially for IoT devices, are resource-restricted. Despite the fact that multiple research papers have been published on incentive systems for IoT devices and edge servers, transaction rejection remains an unsolved issue. Because miners are expected to have a strong motive to record the transaction into a block for a reward, most publications take the success of Blockchain transactions for granted. Edge nodes, being resource-constrained devices, may be unable to expend their energy and participate in the Blockchain system due to the difficulties of recharging.
    \item \textbf{System for Advanced Smart Contracts:} The standard IoT system employs a centralized design in which IoT pieces of equipment communicates information to a cloud network for analysis. The network server, in turn, completes the work and returns to the IoT equipment. The difficulty with the centralized approach is its restricted scalability. This platform is not ideal for instances when devices desire to begin payments with their own interest to others (e.g., a smart healthcare system with automatic payment).
    \item \textbf{Federated Learning for IoT:} The conventional learning technique cannot be employed in the IIoT due to scattered and private datasets. Federated learning is a system in which devices execute training and then transfer the results to the server for data accumulation on a global scale \cite{rahman2023icn}. The advantage of federated learning is that it eliminates the need to migrate a large volume of information from IoT equipment to a cloud server. This is a novel research path, and in order to incorporate Blockchain-assisted FL for resource-bound IoT equipment's, further study is necessary \cite{khan2021federated, kundu2024federated}.
\end{itemize}}

In addition, IIoT, being a relatively new area of exploration, is facing several challenges with the increased level of applications.

\textbf{Real-Time Optimization Issue for IIoT:}~
The implementation of IoT in the industrial sector means increased usage of interconnected devices and a massive amount of data generation and transfer. This results in several issues for wireless communication and creates an opportunity for improved interoperability amongst devices. Since IIoT involves a wide area of applications in several environments and serves diverse demands, a single wireless network cannot perform the required tasks in an IoT platform. Connectivity via a high-performing wireless technology for each IoT application is a necessity since it requires a significant number of standards, numerous frequency bands and communication protocols \cite{lethaby2017wireless}. As such advanced technology in wireless communication such as RFID, Wi-Fi-Direct, and 5G is deployed to enhance the ability of IoT. Cloud computing and edge computing to manage complex database-related tasks creates the structure of communication platforms in IoT. 
The use of cloud computing and edge computing in IIoT is discussed briefly. Systems that are capable of solving real-world problems within specified time constraints require the development of programs based on timing characteristics. IIoT further requires the development of embedded systems with integrated micro-controllers. A real-time embedded system plays a key role in optimizing human-machine interference. An analysis of challenges and future directions in the optimization issue of deploying IIoT is discussed in \cite{nguyen2021real}.\vspace{2mm}

\textbf{Cybersecurity Gaps in IIoT:}~
The ever-expanding usage of IIoT in the industry sector means the cybersecurity risk related to these areas is on the rise. The introduction of the new interconnected device into the manufacturing and infrastructure organizations is creating new areas of attack that has the potential to reveal critical operational functions to cyber-attacks leading to severe consequences. These exposures can be critical to the performance of an organization and create considerably damaging incidents \cite{creese2020future}. As such new technological advancements and cybersecurity systems need to be implemented to solve the technical challenges in an operational system. This means further work is also needed to improve and process regulations and design incentive models to generate the required cybersecurity practices. The characteristics of industrial IoT, which has a potential impact related to cybersecurity, are discussed in \cite{axon2021emerging}. Future research work to understand the nature of systematic risk involving cybersecurity is required to develop tools for preventing and responding to those events. Also, the policies and practices to manage the IIoT from malicious threats need further study.\vspace{2mm}

\textbf{Energy-Efficiency Issue in IIoT:}~
In order to assess the lifetime of IIoT systems and keep track of energy consumption made by sensors, devices, and machines in IIoT, it is important to realize the significance of energy efficiency in this sector. Energy-efficient communication networks and computational models in IIoT systems can considerably reduce the carbon footprint. The sensing and communication tasks in IIoT consume a tremendous amount of energy and potentially impact the life of a smart grid system, as discussed in \cite{li2017smart}. Energy efficiency issue in IIoT systems also impacts the manufacturing and mining industry. A comparison study is presented in \cite{mao2021energy} to understand the impact of energy efficiency issues in IIoT systems based on communication and computational methods. This area requires significant attention to promote green and efficient IIoT systems.

\textbf{Computational models for Smart IIoT:}~
It is essential to perform an analysis to determine the efficacy of several computational models for IIoT prior to the implementation of a particular approach for usability. Significant research into the proposal of innovative computational models and strategies for IIoT means it is necessary to classify the existing algorithm for the proper selection. Researchers performed a thorough analysis and presented summarized comparison models, including the state-of-the-art technologies of the IIoT dealing with implementation models, major developments, and potential challenges. Industrial IoT supports several technologies such as cloud computing, fog computing, edge processing, distributed computing, and cloud figuring \cite{ying2021overview}. The research depicts a survey of the research community to enhance the pattern of proposed algorithms for future research based on the process, development, and monitoring of IIoT applications.

\subsection{\textcolor{black}{Future Opportunities}}
\subsubsection{\textcolor{black}{Role of AI in Cybersecurity}}
\textcolor{black}{Artificial intelligence is built on a self-learning process that may improve notions of comprehending any circumstance using data. As part of the current smart scenario, the creation of data has increased dramatically due to the fast growth of IoT-based systems. The cyber system has to be powerful and safe in order to protect this data. The growing use of IIoT in business means that the cybersecurity risk associated with these sectors is increasing. The introduction of the new networked device into industrial and infrastructure companies has the potential to expose essential operational activities to cyber-attacks, resulting in serious repercussions. To address the technical issues in an operational system, new technology breakthroughs and cybersecurity solutions must be deployed. This indicates that more work is needed to strengthen and process rules as well as create incentive models to encourage the adoption of the necessary cybersecurity procedures. DOS (includes spam), R2L (get local access through packets), user to root (accessing host's rights), and probing (by system access) are some of the most common cyber threats. The scope of AI in the cyber security enhancement system is in the initial stage. Though new but the opportunities are high because of the nature of AI to secure the system.}

\begin{itemize}
    \item \textcolor{black}{ \textbf{False Positives Removal:} In the networking world, the machines needs to perform some important analysis daily, such as traffic analysis, access grant or deny based on the criteria set, and abnormalities detection. In this case, the false-positive results would create a mess for the security system. However, by leveraging AI, the removal of this scenario is possible \cite{van2022initial}.}
    
    \item \textcolor{black}{\textbf{Predicting Abnormalities:} AI has the advantage of being able to learn from the past. With the advancement of internet usage, cyber assaults are growing more complicated; therefore, adequate monitoring and identification of attacks have become a vital element of the industry. In this regard, the AI can forecast the harmful action at the earliest stage with the rationale of the assault. As a result, by detecting the assault early on, the attack mitigation technique will be far more successful in the future \cite{alavizadeh2021markov}.}
    
    \item \textcolor{black}{\textbf{Providing Protection:} In science, AI is compared to a white blood cell in the human body since it does not shut down the entire system in the event of a defect but rather locates and corrects the problem. As a result, the system is protected from failure by evaluating its patterns. As a result, AI can effectively aid in strengthening cyber system security. }
\end{itemize}

\textcolor{black}{However, compared to conventional security methods, the AI-based method will be faster and more convenient. By comprising two methods that are expert systems and neural networks, the AI-based methods detect and mitigate attacks efficiently \cite{sarker2021ai}. }

\subsubsection{\textcolor{black}{Ensuring Security through AI and Blockchain}}

\textcolor{black}{With the growth of data and the usage of the internet, the security of digital data systems must improve. User-sensitive data, in particular, must be handled with greater caution. With the danger to the security system, however, cybersecurity advancements bring new ways to enhance security systems. Cybersecurity is the most pressing problem for every corporation or government in today's digital era because a database attack might expose sensitive information such as passwords or bank account numbers. The Blockchain offers a cryptographic solution for data protection. This allows personal information to be stored outside of the parties' databases. Data is controlled by persons who are data owners, and only metadata is maintained on Blockchain. Zero-knowledge proofs employ cryptographic algorithms to verify the accuracy of propositions without resealing data. Again, AI-based algorithms with Blockchain integration help the system detect if it is under assault or at risk of attack by continuously monitoring blocks in the Blockchain's chain. As a result, the whole system's trust improves, potentially increasing the system's security. The role of AI with Blockchain in the cybersystem are as follows:}

\begin{itemize}
    \item AI-based algorithms identify harmful assaults in the system to increase system security. Additionally, the integration of AI and BC has strong cryptography capabilities making a robust system.
    
    \item Users' security is ensured by smart contracts that compel the formation of dataset permissions. Again, this connection aids the user in deciding whether or not to disclose data.
    
    \item Improves data security through proper data validation and accepts only permissioned data through proper monitorization.
    
\end{itemize}

\subsubsection{\textcolor{black}{Concrete Contribution of Blockchain in Smart Industrial IoT}}
\textcolor{black}{IoT is at the heart of Industry 4.0, and it confronts several issues, including third-party assaults on the IoT ecosystem, data openness, and providing security and privacy to Industry 4.0 applications. Some new technology has been drawn to overcome the limitations of IR 4.0, according to researchers. Blockchain technology, for example, protects data from outsiders by forming a secure chain-like structure.}
\begin{itemize}
 \item As Blockchain keeps information across a couple of computers that form a network, data tampering by attackers becomes more difficult. 
\item The Blockchain mechanism's smart contracts allow for the formation of legitimate and verifiable communication throughout the network. 
\item For IoT sensors, Blockchain might be a viable choice for providing an encryption method as well as a secure environment for execution and identification.
\item Blockchain enables for legitimate boot, counter, firmware, and data encryption for base stations.
\item A stable way of communication is allowed only in the top layers for the public key system and cloud architecture Blockchain
\item By adopting the Blockchain method, decentralization allows for the authentication of centralized structures to decentralized administration in a cost-effective manner.
\item Blockchain provides services such as a traceable and trustworthy means of data transmission by offering an open and trusted medium to ensure the transparency and traceability of data shared between users of the automated sector.

\end{itemize}

\vspace{2mm}
\subsubsection{Security for Industry 4.0 through SDN}
Data plays a vital role in effective communication in this digitalized age. Thus, by offering a stable and secured data transmission system, modern technologies can perform their relevant tasks without interference. 
In industry 4.0, a massive amount of devices are interconnected and continuously transfer data from one device to another. Thus, the system may be compromised by intruders due to the heterogeneity of devices in Industry 4.0 applications. Additionally, the possibility of attacks has escalated with the industrial revolution. Injection attack, Side-channel attack, Time delay attack, False logic attack, Stuxnet attack, Deception attack, DOS attacks, Zero-Day Attacks, Application protocol attack, Fake Location Injection, False Routing Information attack, Man in middle attack, Scan attack, Eavesdropping attack are some of the attacks identified that impact the security of industry 4.0 \cite{rahman2021sdn, jamai2020security}. Moreover, the software-define network divides the total network using its data plane, control plane, and application plane. To maintain a proper organization of the workflow, the controller and switches perform the overall operation. Here, the switches are responsible for moving the data here, and the controller, which is known to be the network brain, controls the route and other related functionalities. Thus, the connection between Industry 4.0 and the SDN environment is made possible with the aid of switches and controllers. From the above-mentioned section, it is clear that SDN is a centralized program. This is centralized as the complete network control process is carried out by a single controller that may have multiple controllers \cite{rahman2023impacts}. \vspace{1mm} 

Furthermore, this nature is useful to ensure security since it can respond quickly to attacks that are destructive to the network, thus providing a firewall for the devices that serve in Industry 4.0. If the threat is immediately identified, the road to minimizing threats is enormously easier. However, SDN ensures security without violating the functionality of the system. By fusing advanced filtering techniques, combining the network switch and firewalls into a robust security concept, and synchronizing security decisions across different layers, SDN provides enhanced security to industry 4.0 applications \cite{rahman2022integration}. Again, the hardware is used to connect devices in conventional systems; on the other hand, the SDN transfers data quite effectively using the software. Being a software system, the data transfer rates, the route of the data, and the detection of attacks in the network are easily maintained and detected, which ensures the network's security. Thus, this method removes the limitations of traditional hardware-connected systems. As this procedure is software-based, it is easy to make a change if required. In addition, SDN maintains a flow table where the information about the connections and the associated properties of the network. The controller of this network decides where to send data based on user requests. By encompassing a model that can identify if the request is from hackers or from a respected user, then the rate of several attacks can be mitigated.

\subsubsection{Merging Industry 4.0 and 5G Network}
From the combining perspective, the combination of Industry 4.0 and 5G has a great impact on the advancement of the industrial revolution. In \cite{rao2018impact}, the authors studied the technologies that can have a great effect on the growth of IR 4.0 applications, where 5G is considered to be one of them. In the smart industry, effective communication between the cloud and the factory is essential. In this case, the 5G network can serve the purpose of wireless communication within the sensors and other parts of the industry \cite{wang20205g}. Another similar research, \cite{perakovic2020development}, has pointed out some challenges regarding the manufacturing process while showing the impact of 5G in the industry's manufacturing module. As the internet is opening new directions to technological advancement, the next-generation internet based on 5G and above can be an effective solution to many challenges faced by the smart industry, smart city, and so on. In addition, for the development of Industry 4.0, these networking schemes also need to be considered \cite{kim20202020, rahman2024blocksd}.

\subsubsection{Collaborative Advantages of SDN, BC, and 5G Services}
The aim of moving toward a 5th generation mobile network is to build one network with higher data rates but lower latency, as well as wide-area coverage, \cite{fourati2021survey}. To develop a common platform that allows for faster data transmission and processing. As people are getting busier in the modern era, and they need to travel from one location to another. They need a networking device that does not disconnect and wants to establish a link as quickly as possible. 5G meets this requirement by maintaining a fast link even when switching from one cell to another. However, switching from one cell to another takes time, which is a contradiction in the 5G networking scheme \cite{haile2021end}. Furthermore, the authentication mechanism needs attention because if it is not done properly, it can cause user data to be harmed, data and history to be lost, and data to be tempered, all of which raises the question of user data protection\cite{ma2020apcn,zhou2019privacy,9290627}. Moreover, as it offers speed connection while changing cells when the user randomly changes their place or their cell, the management and maintenance become much more difficult. Thus there is a need for a method that can efficiently handle the issues of the network to make it suitable for use holding its objectives. Being heterogeneous, privacy protection requires some intelligent measure like Blockchain or SDN or some other methods that have recently advanced and are handling a lot of issues in the wireless community \cite{feng2021efficient,datsika2021sdn}. The benefit of combining SDN in the 5G networking arena is mainly the separation of control and data plane. The advanced feature provided by SDN (flexibility) will be able to fulfill the enormous demand of the 5G network, additionally maintaining the service quality and providing a better pathway for communication (M2M, H2H) \cite{kim2021location}. Blockchain technology will help the network overcome the security challenge with the help of a cryptographic solution. With a decentralized server, a better and more secure network is possible among the users of the 5G network \cite{serrano2021blockchain}. No third-party intervention and secured and open access data transfer a reliable, transparent, and secured connection is maintained \cite{tan2021blockchain}. Thus, the combination of SDN, Blockchain, and 5G will enable the wireless communication system to be a better and more secure solution for connection establishments.

\subsubsection{\textcolor{black}{Impact of Smart Industrial IoT in Smart Agriculture}}
\textcolor{black}{Industrial IoT is essentially the internet of things that aids businesses in monitoring the instruments that execute certain tasks. Many industries have improved their security, production, and efficiency by embracing IIoT. Farmers can better monitor their crops with the use of smart IIoT in the agricultural area. The essential principle here is to obtain visibility of the crop farming process through the effective application of smart sensors. The sensor data capture weather and water level data. The data was stored for future reference in case an error occurred. Smart sensor readings also assist farmers in growing crops without the use of pesticides and with the right use of natural resources. Environmental records also assist specialists in forecasting future agricultural conditions. Smart sensors, on the other hand, may be used to manage the greenhouse weather. Some of the main applications of Smart IIoT in the agriculture fields are--}

\begin{itemize}
    \item \textcolor{black}{\textbf{Nurture Soil:} The soil is the main food for crops. Without proper nurture or preparation of soil, crops will never grow well. By leveraging smart sensors like humidity and water level sensors, the actual soil data is achievable, and by analyzing their sensor data, the soil's level of nutrition is found. These sensor data provide insights of soil moisture level, water holding capacity, nutrition in the soil and deficiencies as well. Robot with GPS system enables the user to get data of the global map by which the farmers get the knowledge of where the water level is nearby \cite{kumar2021design}}
    \item \textcolor{black}{\textbf{Fertilization:} Soil fertilization is mostly influenced by the soil's water level. However, the water resource is depleting day by day. As a result, appropriate water management is another critical responsibility in making the soil productive. For this reason, IIoT methods such as sprinkler irrigation or drip irrigation can be beneficial \cite{garcia2020iot}.}
    \item \textcolor{black}{\textbf{Disease Detection and Proper Management:} Drones and small cameras have been developed to help professionals record real-time plant data. Because IoT devices generate data in real-time and on a regular basis, system control is more accurate and quick, with no need for human intervention. Drones can also be used to do real-time field monitoring. Furthermore, accurate diagnosis of plant disease is achievable thanks to IIoT-based intelligent mechanisms. Through the use of IIoT-based sensors and systems, accurate detection and identification of appropriate pesticides are also achievable \cite{kavitha2021industrial}. }
    
\end{itemize}

\subsubsection{\textcolor{black}{Smart Healthcare Development through BC-AI Techniques}}
\textcolor{black}{Integration of numerous smart technologies and data transfer has expanded dramatically in a smart healthcare system, implying the significance of safeguarding healthcare data because it contains sensitive user information. AI (OCR technology) aids the staff in retrieving stored data in a short amount of time while ensuring Blockchain-based data security \cite{singh2020blockchain, 9642537}. In a smart healthcare system, the production of data is vast, and the total system depends on this data to make everything automated. However, the data in this sector is sensitive and needs high-level security protocols to be followed. Blockchain is a novel and promising technology that is mostly employed in situations where centralization is inappropriate, and privacy is crucial \cite{jabarulla2021blockchain}. The data is saved in blocks that are spread within a network using the fundamentals of Blockchain, which are decentralization and cryptographic solutions. The user can access this data with the appropriate key, and if the key does not match, the system will be protected by raising the alarm through the Blockchain. Smart contracts are also used to guarantee extra transparency amongst system users. The correct monitoring of each activity is also feasible by using the transaction's identifying number. Integrating AI in the healthcare sector with secured Blockchain enables methods the intelligence of the system improves. The use of robots in surgery has gained favor as intelligent agent-based robotics has advanced. However, the usage of robots in this capacity creates security issues. AI method is based on algorithms like XGBoost plus Blockchain integration, where the Blockchain-based approach complements the AI-based EHR data storage mechanism, in which AI maintains system security and Blockchain cryptographically manages EHR data. It assists healthcare professionals in making quick decisions in very critical situations by properly analyzing the patients' reports through algorithms \cite{rahman2024internet}. }
\begin{itemize}
    \item \textcolor{black}{Blockchain allows for improved information exchange between multiple healthcare organization platforms in a safe manner.}
    \item \textcolor{black}{Patients' sensitive records are securely saved and maintained using a Blockchain-based data storage system for future reference by patients and healthcare providers. Thus remote monitoring of patients is also possible \cite{singh2020blockchain}}
    \item \textcolor{black}{With the decentralized database of healthcare information, Blockchain has aided in the study and development of healthcare-based systems and therapy.}
    \item \textcolor{black}{AI can help with security improvement and automation. Healthcare practitioners benefit from AI's automated monitoring system.}
    \item \textcolor{black}{The sophisticated AI-based algorithms aid in the analysis of health-related pictures such as X-ray images, MRI images, ultrasound images, and so on for early illness identification and prediction.}
    \item \textcolor{black}{In the event of a pandemic, AI can also help the pharmaceutical industry research new medications.}
    \item \textcolor{black}{The use of robots in surgery has gained favour as intelligent agent-based robotics has advanced. However, the usage of robots in this capacity creates security issues. AI method is based on  algorithms like XGBoost plus Blockchain integration where the Blockchain-based approach complements the AI-based EHR data storage mechanism, in which AI maintains system security and Blockchain cryptographically manages EHR data \cite{tagde2021blockchain}.}
    
\end{itemize}

\subsubsection{\textcolor{black}{Role of AI and Smart IIoT to Mitigate the 6G challenges}}
\textcolor{black}{The wireless communication system is moving toward automation, which can be accomplished via AI-based smart technologies. However, with many advancements come new obstacles. The most important of them is intrusion detection and mitigation in a 6G system. The use of AI in maintaining privacy and security.
Sub-networks in 6G, which can be thought of as an extension of local 5G networks outside vertical domains, may derive assistance from training-based security solutions both within and across sub-networks. Other sub-networks' behavior can be captured, and malicious traffic can be detected using ML-based algorithms placed at the perimeter. Massive data transfers between sub-networks may be ineffective because these networks are typically self-contained. For communication efficiency, a sub-network can solely communicate the learned security information with another. A second sub-network can take the collaborative data, input it into its machine learning models, and apply dynamic policies to other networks' hostile traffic.\newline
In contrast to current centralized cloud-based AI systems, 6G will rely heavily on edge intelligence. In the vast device and data regime \cite{ma2020safeguarding}, the dispersed nature facilitates the implementation of edge-based FL for network safety, ensuring communication efficiency. 6G design envisions connected intelligence and employs artificial intelligence at many levels of the network structure \cite{letaief2019roadmap}. At the cellular level, artificial intelligence has the ability to prevent DoS attacks on cloud networks. The ability of a device to enable multiple connections in a mesh network enables numerous base stations to analyze the behavior of equipment using AI classification programs and coordinately make decisions on the verification with the use of weighted average techniques, as described in \cite{chkirbene2020weighted}.
AI-powered predictive analytics can identify attacks, such as 51 percent attacks on Blockchain, before they happen. A quantum computer could put asymmetric key cryptography in jeopardy. They can, however, give exponential speedups for AI/ML systems, allowing them to complete previously impossible jobs considerably faster. In 6G, multi-connectivity mesh networks with tiny cells enable devices to communicate simultaneously via numerous base stations. Edge-based machine learning models might be used to dynamically discover privacy-preserving routes, rate them, and enable devices to transfer data via those routes based on the ranking. In contrast to cloud-based centralized learning, federated learning keeps data close to the user, enhancing data and location confidentiality. The 6G subnetwork level AI maintains confidentiality within the subnetwork and only shares learned data with the outside world to reduce security threats. AI-based services privacy-preserving policy updates \cite{liyanage20185g} have the ability to provide wholly automated 6G networks while maintaining privacy.}

\section{Conclusion}
\label{sec:concl}
\textcolor{black}{In this survey, the authors provided a piece of state-of-the-art information regarding Blockchain, AI, and smart IIoT on the basis of the current world demand. This article offers excellent integration of these (BC-AI-IIoT) techniques. Moreover, this survey efficiently discusses different issues--security, privacy, confidentiality, and so on based on these technologies. Then, it also presents the various tabular analyses regarding the vast areas. A significant number of solutions, tools, and techniques have been extensively analyzed in this survey.
Additionally, some open issues and opportunities have been discussed in this research. Although this research can provide current and innovative information on the basis of BC, AI, and IIoT technologies, this article still needs to cover extra information that can properly mitigate the various problems and challenges in the different domains. In the future, the authors suggest some taxonomies with proper groups and apply these technologies in the other fields, such as smart cities, innovative healthcare, smart vehicle management, etc.}




    


\clearpage
\balance
\bibliographystyle{elsarticle-num-names}

\bibliography{sample.bib}


\end{document}